\documentclass{aa}

\usepackage{graphicx}
\usepackage{txfonts}

\newcommand{\be}{\begin{equation}}
\newcommand{\ee}{\end{equation}}
\newcommand{\ba}{\begin{eqnarray}}
\newcommand{\ea}{\end{eqnarray}}
\newcommand{\bc}{\begin{center}}
\newcommand{\ec}{\end{center}}

\begin{document}

\title{High energy $\gamma$-ray emission from the \\ starburst nucleus of NGC~253}
\author{Eva Domingo-Santamar\'{\i}a$^{1}$ \& Diego F. Torres$^{2,1}$}

\institute{Institut de F\'{\i}sica d'Altes Energies (IFAE),
Edifici C-n, Campus UAB, 08193 Bellaterra, Spain.
\and Lawrence Livermore National Laboratory, 7000
East Avenue, L-413, Livermore, CA 94550, USA.
}

\authorrunning {Eva
Domingo-Santamar\'{\i}a \& Diego F. Torres}

\titlerunning{NGC~253}

\abstract{The high density medium that characterizes the central
regions of starburst galaxies and its power to accelerate
particles up to relativistic energies make these objects good
candidates as $\gamma$-rays sources. In this paper, a
self-consistent model of the multifrequency emission of the
starburst galaxy NGC~253, from radio to $\gamma$-rays, is
presented. The model is in agreement with all current measurements
and provides predictions for the high energy behavior of the
NGC~253 central region. Prospects for observations with the HESS
array { (and comparison with their recently obtained data)} and
GLAST satellite are especially discussed. }

\maketitle

\section{Introduction}

Starburst galaxies (and in general star forming regions) are
expected to be detected as $\gamma$-ray sources. They have both a
large amount of target material and, due to the presence of
supernova remnants and the powerful stellar winds of their
numerous young stars, multiple shocks where to accelerate
particles up to relativistic energies. Pion decay production of
$\gamma$-rays, usually dominant under such conditions, is thought
to produce significant $\gamma$-ray fluxes.

Approximately 90\% of the high-energy $\gamma$-ray luminosity of
the Milky Way ($\sim$1.3 $\times 10^6$ L$_\odot$, Strong,
Moskalenko, \& Reimer 2000) is diffuse emission from cosmic ray
interactions with interstellar gas and photons (Hunter et al.
1997).  However,  the LMC is the only external galaxy that has
been detected in its diffuse $\gamma$-ray emission (Sreekumar et
al. 1992), a fact explained by the isotropic flux dilution by
distance. At 1 Mpc, for example, the flux of the Milky Way would
approximately be 2.5 $\times 10^{-8}$ photons cm$^{-2}$ s$^{-1}$
($>$100 MeV), what is below the sensitivity achieved up to now in
the relevant energy domain. The Energetic $\gamma$-ray Experiment,
EGRET, did not detect any starburst. Upper limits were imposed for
M82,  $F(E>100 {\rm MeV}) < 4.4 \times 10^{-8}$ photons cm$^{-2}$
s$^{-1}$, and NGC~253,  $F(E>100 {\rm MeV}) < 3.4 \times 10^{-8}$
photons cm$^{-2}$ s$^{-1}$ (Blom et al. 1999), the two nearest
starbursts, as well as upon many luminous infrared galaxies, for
which similar constraints were found (Torres et al. 2004). These
upper limits are barely above the theoretical predictions of
models for the $\gamma$-ray emission of galaxies, constructed with
different levels of detail (see Torres 2004b for a review).
Starbursts and luminous infrared galaxies are expected to
compensate the flux dilution produced by their relatively larger
distance to Earth with their higher cosmic ray flux, and become
sources for the next generation of instruments measuring photons
in the 100 MeV -- 10 GeV regime, like the Gamma-ray Large Area
Telescope, GLAST (e.g., V\"olk, Aharonian \& Breitschwerdt 1996,
Paglione et al. 1996, Blom et al. 1999, Torres et al. 2004, Torres
2004).

In this paper, we analyze one such galaxy, the well studied
starburst NGC~253. Herein, we present a self-consistent
multiwavelength modelling of the central region of the galaxy,
taking into account the latest measurements of densities,
supernova explosion rate, radio emission, and molecular mass,
among other physical parameters that we use as input for our
scenario.

\section{CANGAROO observations of NGC~253 }

Recently, the CANGAROO collaboration reported the detection of
NGC~253 at TeV $\gamma$-ray energies (an 11$\sigma$ confidence
level claim), observed during a period of two years in 2000 and
2001 by about 150 hours (Itoh et al. 2002, 2003). Their measured
differential flux was fitted by Itoh et al. (2003) with a power
law and an exponential cutoff, obtaining
\begin{eqnarray*}
\frac{dF}{dE}=(2.85\pm 0.71)\times 10^{-12}\, (E/{\rm 1
TeV})^{(-3.85\pm 0.46)} ~{\rm cm^{-2}s^{-1}TeV^{-1}},
\end{eqnarray*}
and
\begin{eqnarray*}
\frac{dF}{dE}=a e^{\sqrt{E_0}/b} (E/E_0)^{-1.5}e^{-\sqrt{E}/b}
~{\rm cm^{-2}s^{-1}TeV^{-1}},
\label{cangaroo}
\end{eqnarray*}
with $ a=6 \times 10^{-5}~{\rm
cm^{-2}s^{-1}TeV^{-1}},~E_0=0.0002~{\rm TeV}, $ and $ b=0.25 \pm
0.01~\sqrt{\rm TeV}\label{eq3}. $ Both parameterizations are
sensible reproductions of the observational data, although the
former is clearly preferred for simplicity upon the light of an
equally good fit. The flux uncertainty are the square root of the
quadratic sum of the statistical and systematic errors. Note that
the slope of the power law spectrum is very uncertain, but steep.
Indeed, an extrapolation of this power-law spectrum to lower
energies violates the measured upper limits in the GeV regime. The
CANGAROO collaboration suggested  a turn-over below the TeV region
and proposed the second spectral form. They have also claimed that
the emission at the highest energies is inconsistent with it being
produced in a point like source, and proposed for it an inverse
Compton origin in a kpc-scale $\gamma$-ray halo.

The HESS array have observed NGC~253 {(see below)}. In several
other observations of sources that have previously been targets
for CANGAROO, the HESS collaboration have presented results in
clear contradiction with the former CANGAROO reports. This is most
notably the case for SN 1006 (Aharonian et al. 2005a), PSR
B1706-44 (Aharonian et al. 2005b), and to some extent also for the
supernova remnant RX J1713-3857 (Aharonian et al. 2004a), and the
Galactic Center (Aharonian et al. 2004b). This may suggest some
kind of systematic difference in the treatment of both sets of
observational data. Such systematic effect should explain why
CANGAROO spectra are steeper and their measured fluxes are one
order of magnitude higher than the upper limits or measurements
obtained with HESS. The CANGAROO collaboration is now calibrating
their stereo system, and will be re-observing these problematic
cases within a year or so (R. Enomoto 2005, private
communication).

In what follows we focus on producing a detailed multiwavelength
theoretical model for the central region of NGC~253, irrespective
of CANGAROO measurements (i.e., we will not try to reproduce their
spectrum, but we will derive predictions of fluxes based on a set
of well-founded assumptions). The aim is to see whether a model
based on observations at all wavelengths and first principles
would --while being consistent with multiwavelength testing--
predict that the central region of NGC~253 alone can produce a
sufficiently high $\gamma$-ray flux so as to be detected by the
current ground-based Cerenkov telescopes {and future MeV--GeV
satellites}. The central region of the galaxy would look like a
point like source for the field of view and angular resolution of
imaging Cerenkov telescopes. Then, we shall explore if we would
rather theoretically expect a HESS non-confirmation of CANGAROO
results regarding perhaps both, the flux and the extension.

\section{Phenomenology of the central region of NGC~253}

A wealth of new multiwavelength data was obtained for NGC~253 during the last
decade (after the previous modelling by Paglione et al. 1996, which
did not include photons energies above 100 GeV, see below).
%
%
It is located at a distance of $\sim 2.5$ Mpc (Turner and Ho 1985,
Maurbersger et al. 1996) and it is a nearly edge-on (inclination
~78$^o$, Pence 1981) barred Sc galaxy. The continuum spectrum of
NGC~253 peaks in the FIR at about 100 $\mu$m, with a luminosity of $4
\times 10^{10}$ L$_\odot$ (Telesco \& Harper 1980, Rice et al.
1988, Melo et al. 2002). The FIR luminosity is at least a factor
of 2 larger than that of our own Galaxy (Cox \& Mezger 1989,
Dudley \& Wynn-Williams 1999), and it mainly (about half of it
according to Melo et al.'s (2002) 1 arcmin resolution ISOPHOT
observations) comes from the central nucleus. IR emission can be
understood as cold ($T \sim 50 K$) dust reprocessing of stellar
photon fields.


When observed at 1 pc resolution, at least 64 individual compact
radio sources have been detected within the central 200 pc of the
galaxy (Ulvestad \& Antonucci 1997), and roughly 15 of them are
within the central arcsec of the strongest radio source,
considered to be either a buried active nucleus or a very compact
SNR. Of the strongest 17 sources, about half have flat spectra and
half have steep spectra. This indicates that perhaps half of the
individual radio sources are dominated by thermal emission from H
II regions, and half are optically thin synchrotron sources,
presumably SNRs. There is no compelling evidence for any sort of
variability in any of the compact sources over an 8 yr time
baseline.
%
%
The most powerful flat-spectrum central radio source is clearly
resolved in the study of Ulvestad and Antonucci (1997) and appears
to be larger than the R136 cluster located in 30 Doradus,
containing about 10$^5$ M$_\odot$ in stars and 600 M$_\odot$ in
ionized gas. The age was estimated to be less than 4 $\times $
10$^6$ yr.
The region surrounding the central ~200 pc has also been observed
with subarcsec resolution and 22 additional radio sources stronger
than 0.4 mJy were detected within 2kpc of the galaxy nucleus
(Ulvestad 2000). The region outside the central starburst
may account for about 20\% of the star formation of NGC~253, is
subject to a supernova explosion rate well below $0.1$ yr$^{-1}$,
and has an average gas density in the range 20--200 cm$^{-3}$,
much less than in the most active nuclear region of NGC~253
(Ulvestad 2000).

%
%

Carilli (1996) presented low frequency radio continuum
observations of the nucleus at high spatial resolution. Free-free
absorption was claimed to be the mechanism producing a flattening
of the synchrotron curve at low energies, with a turnover
frequency located between 10$^{8.5}$ and 10$^9$ Hz. The emission
measures needed for this turnover to happen, for temperatures in
the order of 10$^4$ K, is at least 10$^5$ pc cm$^{-6}$. Tingay
(2004) observed NGC~253 using the Australian Long Baseline Array
and provided fits with free-free absorption models for the radio
spectrum of six sources. He concluded that the free-free opacity
in the central region has to be in the range of 1 to 4 at 1.4 GHz,
implying emission measures of a few times 10$^6$ pc cm$^{-6}$ in
this particular directions, again for temperatures of the order of
10$^4$ K.


As shown by infrared, millimeter, and centimeter observations, the
200 pc central region dominates the current star formation in
NGC~253, and is considered as the starburst central nucleus (e.g.,
Ulvestad and Antonucci 1997, Ulvestad 2000). Centimeter imaging of
this inner starburst, and the limits on variability of radio
sources, indicates a supernova rate less than 0.3 yr$^{-1}$
(Ulvestad \& Antonucci 1997), which is consistent with results
ranging from 0.1 to 0.3 yr$^{-1}$ inferred from models of the
infrared emission of the entire galaxy (Rieke et al. 1980; Rieke,
Lebofsky \& Walker 1988, Forbes et al. 1993). Van Buren and
Greenhouse (1994) developed, starting from Chevalier's (1982)
model for radio emission from supernovae blast waves expanding
into the ejecta of their precursor stars, a direct relationship
between the FIR luminosity and the rate of supernova explosions.
The result is ${\cal R} = 2.3 \times 10^{-12} L_{\rm FIR}/L_\odot$
yr$^{-1}$, which is in agreement, within uncertainties, with the
previous estimates. The star formation rate at the central region
has been computed from IR observations, resulting in 3.5 M$_\odot$
yr$^{-1}$, and represents about 70\% of the total star formation
rate measured for NGC~253 (Melo et al. 2002). When compared with
Local Group Galaxies, the supernova rate in NGC~253 is one order
of magnitude larger than that of M31, the largest of the Local
Group (Pavlidou and Fields 2001).


Paglione et al. (2004) obtained high resolution (5".2 $\times$
5".2) interferometric observations of the CO line $J=1 \rightarrow
0$ in order to study the structure and kinematics of the molecular
gas in the central nucleus. This study enhances that of Sorai et
al. (2000), which, although done with less angular resolution,
obtained compatible results. The general morphology of the CO map
is consistent with other high resolution studies. It shows an
extended ridge of emission aligned with an infrared-bright bar and
a central group of bright clouds aligned with the major axis of
the galaxy, orbiting the radio nucleus in a possible ring. The
central clouds move around the radio nucleus as a solid body,
similar to the distribution of the radio sources, central HCN
clouds, and central near-infrared emission (Paglione et al. 1995,
1997; Ulvestad \& Antonucci 1997).
%
%
Much of the molecular gas in NGC~253 appears to be highly excited
(Wild et al. 1992; Mao et al. 2000; Ward et al. 2003).
Observations of $J=4 \rightarrow 3$ and $J=6 \rightarrow 5$
transitions of CO as well as HCN lines suggest the existence of
localized spots with values of densities in excess of 10$^4$
cm$^{-3}$ (Israel \& Baas 2002, Paglione, Jackson, \& Ishizuki
1997, Paglione, Tosaki \& Jackson 1995, Harris et al. 1991).
Bradford et al. (2003) have reported CO $J=7 \rightarrow 6$
observations and also find that the bulk of molecular gas in the
central 180 pc of the galaxy is highly excited and at a
temperature of about 120 K. They concluded that the best mechanism
for heating the gas is cosmic ray bombardment over the
gas residing in clouds, with
density about 4.5 $\times 10^4$ cm$^{-3}$.

Current estimates of the gas mass in the central 20'' -- 50'' ($<
600$ pc) region range from 2.5 $\times 10^{7}$ M$_\odot$
(Harrison, Henkel \& Russell 1999) to 4.8 $\times 10^{8}$
M$_\odot$ (Houghton et al. 1997), see Bradford et al. (2003),
Sorai et al. 2000, and Engelbracht et al. (1998) for discussions.
For example, using the standard CO to gas mass conversion, the
central molecular mass was estimated as 1.8 $\times 10^{8}$
M$_\odot$ (Mauersberger et al. 1996). It would be factor of $\sim
3$ lower if such is the correction to the conversion factor in
starburst regions which are better described as a filled
intercloud medium, as in the case of ULIRGs, instead of a
collection of separate large molecular clouds, see Solomon et al.
(1997), Downes \& Solomon (1998), and Bryant \& Scoville (1999) for
discussions. Thus we will assume in agreement with the mentioned
measurements that within the central 200 pc, a disk of 70 pc
height has $\sim$ 2 $\times 10^{7}$ M$_\odot$ uniformly
distributed, with a density of $\sim 600$ cm$^{-3}$. Additional
target gas mass with an average density of $\sim$50 cm$^{-3}$ is
assumed to populate the central kpc outside the innermost region,
but subject to a smaller supernova explosion rate $\sim 0.01$
yr$^{-1}$, 10\% of that found in the most powerful nucleus
(Ulvestad 2000).


The central region of this starburst is packed with massive stars.
Watson et al. (1996) have discovered four young globular clusters
near the center of NGC~253; they alone can account for a mass well
in excess of 1.5$\times 10^6 $M$_{\odot}$ (see also Keto et al.
1999). Assuming that the star formation rate has been continuous
in the central region for the last 10$^9$ yrs, and a Salpeter IMF
for 0.08--100 M$_{\odot}$, Watson et al. (1996) find that the
bolometric luminosity of NGC~253 is consistent with 1.5 $\times
10^8 $M$_{\odot}$ of young stars.
%
%
Physical, morphological, and kinematic evidence for the existence
of a galactic superwind has been found for NGC~253 (e.g., McCarthy
et al. 1987, Heckman et al. 1990, Strickland  et al. 2000, 2002,
Pietsch et al. 2001, Forbes et al. 2000, Weaver et al. 2002,
Sugai, Davies \& Ward 2003). This superwind creates a cavity of
hot ($\sim10^8$ K) gas, with cooling times longer than the typical
expansion timescales. As the cavity expands, a strong shock front
is formed on the contact surface with the cool interstellar
medium. Shock interactions with low and high density clouds can
produce X-ray continuum and optical line emission, respectively,
both of which have been directly observed (McCarthy et al. 1987).
The shock velocity can reach thousands of km s$^{-1}$. This wind
has been proposed as the convector of particles which have been
already accelerated in individual SNRs, to the larger superwind
region, where Fermi processes could upgrade their energy up to
that detected in ultra high energy cosmic rays (Anchordoqui et al.
1999, Anchordoqui et al. 2003, Torres \& Anchordoqui 2004).

\section{Diffuse modelling}

The approach to compute the steady multiwavelength emission from
NGC~253 follows that implemented in ${\cal Q}$-{\sc diffuse},
which we have used with some further improvements (Torres 2004).
The  { steady state particle distribution} is computed within
${\cal Q}$-{\sc diffuse} as the result of an injection
distribution being subject to losses and secondary production in
the ISM. In general, the injection distribution may be defined to
a lesser degree of uncertainty when compared with the steady state
one, since the former can be directly linked to observations. The
injection proton emissivity, following Bell (1978), is assumed to
be a power law in proton kinetic energies, with index $p$, $
Q_{\rm inj}(E_{\rm p,\, kin}) = K ({E_{\rm p,\, kin}}/{\rm
GeV})^{-p}$ , where $K$ is a normalization constant and units are
such that $[Q]$= {\rm GeV}$^{-1}$ {\rm cm}$^{-3}$  {\rm s}$^{-1}$.
The normalization $K$ is obtained from the total power transferred
by supernovae into CRs kinetic energy within a given volume $
\int_{E_{\rm p,\,kin,\, min}}^{E_{\rm p,\, kin,\, max}} Q_{\rm
inj}(E_{\rm p,\, kin}) E_{\rm p,\, kin} dE_{\rm p,\, kin} = -K
{E_{\rm p,\,kin,\,min}^{-p+2}}/(-p+2)
 \equiv { \sum_i \eta_i {\cal P} {\cal R}_i }/ {V}
$ where it was assumed that $p\neq 2$, and used the fact that
$E_{\rm p,\,kin\,min} \ll E_{\rm p,\,kin,max}$ in the second
equality. ${\cal R}_i$ ($\sum_i {\cal R}_i={\cal R}$) is defined
as the rate of supernova explosions in the star forming region
being considered, $V$ being its volume, and $\eta_i$, the
transferred fraction of the supernova explosion power (${\cal P}
\sim 10^{51}$ erg) into CRs (e.g., Torres et al. 2003 and
references therein). The summation over $i$ takes into account
that not all supernovae will transfer the same amount of power
into CRs (alternatively, that not all supernovae will release the
same power). The rate of power transfer is assumed to be in the
range 0.05 $\lesssim \eta_i \lesssim$ 0.15, the average value for
$\eta$ being $\sim 10\%$. {At low energies the distribution of
cosmic rays is
 flatter, e.g., it would be given by equation (6) of Bell (1978),
 correspondingly normalized. We have numerically verified that to neglect
 this difference at low energy
 does not
 produce any important change in the computation of secondaries,
and especially on $\gamma$-rays at the energies of interest.}

The general diffusion-loss equation is given by (see, e.g.,
Longair 1994, p. 279; Ginzburg \& Syrovatskii 1964, p. 296) \be -
D \bigtriangledown ^2 N(E)+\frac{N(E)}{\tau(E)} - \frac{d}{dE}
\left[ b(E) N(E) \right] - Q(E) = - \frac{\partial N(E)}{\partial
t} \label{DL} \ee In this equation, $D$ is the scalar diffusion
coefficient, $Q(E)$ represents the source term appropriate to the
production of particles with energy $E$, $\tau(E)$ stands for the
confinement timescale, $N(E)$ is the distribution of particles
with energies in the range $E$ and $E+dE$ per unit volume, and
$b(E)=-\left( {dE}/{dt} \right)$ is the rate of loss of energy.
The functions $b(E)$, $\tau(E)$, and $Q(E)$ depend on the kind of
particles. In the steady state,  $ {\partial N(E)}/{\partial t}
=0,$ and, under the assumption of a homogeneous distribution of
sources, the spatial dependence is considered to be irrelevant, so
that $ D \bigtriangledown ^2 N(E) =0.$  Eq. (\ref{DL}) can be
solved by using the Green function  \be G(E,{E^\prime})= \frac
{1}{b(E)} \exp \left( -\int_E^{E^\prime} dy \frac{1}{\tau(y) b(y)}
\right), \ee such that for any given source function, or
emissivity, $Q(E)$, the solution is \be N(E) = \int_E^{E_{\rm
max}} d{E^\prime} Q({E^\prime}) G(E,{E^\prime}). \label{SOL-DL}\ee
The confinement timescale will be { determined by several
contributions}. One on hand, we consider the characteristic escape
time in the homogeneous diffusion model (Berezinskii et al. 1990,
p. 50-52 and 78) $ \tau_D= {R^2}/( {2D(E)}) ={\tau_0}/( \beta
(E/{\rm GeV})^{\mu} ), \label{T-P0} $ where $\beta$ is the
velocity of the particle in units of $c$, $R$ is the spatial
extent of the region from where particles diffuse away, and $D(E)$
is the energy-dependent diffusion coefficient, whose dependence is
assumed $\propto E^{\mu}$, with $\mu \sim 0.5$.\footnote{ {We
emphasize that the use of an homogeneous model is an
 assumption, but justified in the compactness of the
 innermost starburst region. We are basically assuming that there is an
 homogeneous distribution of supernovae in the central
 hundreds of pc, what is supported observationally
 (Ulvestad and Antonucci 1997).}} $\tau_0$ is the
characteristic diffusive escape time at $\sim$ 1 GeV.  {$\tau_0$
for NGC253 is not known.
 One can only assume its value and compare it
 with that for other galaxies (e.g. our own Galaxy, or M33,
 Duric et al. 1995);
 the value we choose also parallels that obtained in an earlier
 study of NGC 253 or on M82 (Paglione et al. 1996, Blom et al. 1999).
 We analyze the sensitivity of the model to $\tau_0$ below.}
 On the other, the total escape
timescale will also take into account that particles can be
carried away by the collective effect of stellar winds and
supernovae. $\tau_c$, the convective timescale, is $\sim R / V$,
where $V$ is the collective wind velocity. {For a wind velocity of
300 km/s and a
 radius of about the size of the innermost starburst (see below), this
 timescale is less than a million years (~3$\times 10^{5}$ yr).
 The outflow velocity is not very well known,
however, but minimum reasonable values between 300 and 600 km
s$^{-1}$ have been claimed, and could even reach values of the
order of thousand of km s$^{-1}$ (Strickland et al. 2002).} { Pion
losses (which are catastrophic, since the inelasticity of the
collision is about 50\%) produce a loss timescale $\tau_{\rm
pp}^{-1}=(dE/dt)^{\rm pion}/E$ (see, e.g., Mannheim \&
Schlickeiser 1994), which is similar in magnitude to the
convective timescale and dominates with it the shaping of the
proton spectrum. Thus, in general, for energies higher than the
pion production threshold $ \tau^{-1}(E)= \tau_D^{-1}
 + {\tau_c}^{-1} +
\tau_{\rm pp}^{-1}. \label{T-P} $} For electrons, the total rate
of energy loss considered is given by the sum of that involving
ionization, inverse Compton scattering, bremsstrahlung, and
synchrotron radiation. We have also incorporated adiabatic losses.
Full Klein-Nishina cross section is used while computing photon
emission, and either Thomson or extreme Klein-Nishina
approximations, as needed, are used while computing losses. For
the production of secondary electrons, ${\cal Q}$-{\sc diffuse}
computes knock-on electrons and charged pion processes producing
both electrons and positrons. In the case of $\gamma$-ray photons,
we compute bremsstrahlung, inverse Compton and neutral pion decay
processes. For the latter, an Appendix provides a more detailed
discussion of the different approaches to compute the neutral
pion-induced $\gamma$-ray emissivity. For radio photons, we
compute, using the steady distribution of electrons, the
synchrotron, and free-free emission. Free-free absorption is also
considered in order to reproduce the radio data at low
frequencies. The FIR emission is modelled with a dust emissivity
law given by $\nu^{\sigma}B(\epsilon, T)$, where $\sigma=1.5$ and
$B$ is the Planck function. The computed FIR photon density is
used as a target for inverse Compton process as well as to give
account of losses in the $\gamma$-ray scape. The latter basically
comes from the opacity to $\gamma\gamma$ pair production with the
photon field of the galaxy nucleus. The fact that the dust within
the starburst reprocesses the UV star radiation to the less
energetic infrared photons implies that the opacity to
$\gamma\gamma$ process is significant only at the highest
energies. The opacity to pair production from the interaction of a
$\gamma$-ray photon in the presence of a nucleus of charge $Z$ is
considered too. For further details and relevant formulae see
Torres (2004).

\subsection{Comparison with previous models}

When compared with the previous study on high energy emission from
NGC~253, by Paglione et al. (1996), several methodological and
modelling differences are to be mentioned. Paglione et al.'s
assumed distance, size, gas mass, density, and supernova explosion
rate of the central region are different from those quoted in
Table 1. The former authors modelled, based on earlier data (e.g.,
Canzian et al. 1988) a starburst region at 3.4 Mpc (a factor of
1.36 farther than the one currently adopted), of 325 pc radius
(about 3 times larger than the one adopted here). This region is
larger than what is implied by the current knowledge of the
central starburst, where the supernova explosion rate Paglione et
al. used is actually found and the cosmic ray density is maximally
enhanced. The average density assumed by Paglione et al., 300
cm$^{-3}$, gives a target mass $\sim 2 \times 10^{8}$ M$_\odot$,
which is at the upper end of all current claims for the central
nucleus, or already found as excessive. The target mass of the
innermost region differs from ours by a factor of about 6, ours
being smaller. The fraction of the supernova explosion converted
into cosmic rays (20\% for Paglione et al., a factor of 2 larger
than ours) seems also excessive in regards of the current
measurements of SNR at the highest energies. We have also
considered, especially to test its influence in the total
$\gamma$-ray output, a surrounding disk with a smaller supernova
rate, following the discovery of several SNRs in that region
(Ulvestad 2000). Finally, Paglione et al. (1996) study did not
produce results above 200 GeV. \footnote{To further ease the
comparison, we here note some typos in Paglione et al. (1996)
paper: The factor $b(E)$ should be elevated to the minus one in
their equation (4), as well as the term $\tau_c$ in their equation
(3). The y-axis of their Figure 1 is not the emissivity, but the
emissivity divided by the density; units need to be accordingly
modified, see e.g. Abraham et al. (1966). $E_p$ in their equation
(7) and $x$-axis of Figure 2 and 3 is the kinetic energy, but the
generic $E$ in Equation (1) is the total energy. The $y$-axis of
Figure 2 is in units of cm$^{-3}$ GeV$^{-1}$. }

The ${\cal Q}$-{\sc diffuse} set uses different parameterizations
for pion cross sections as compared with those used by
Marscher and Brown (1978), whose code was the basis of Paglione et al.'s
study. Our computation of neutral pion decay $\gamma$-rays is
discussed in the Appendix. In addition, ${\cal Q}$-{\sc diffuse}
uses the full inverse Compton Klein-Nishina cross section,
computes secondaries (e.g., knock-on electrons) without resorting
to parameterizations which are valid only for Earth-like cosmic
ray (CR) intensities, fixes the photon target for Compton
scattering starting from modelling of the observations in the FIR,
and considers opacities to $\gamma$-ray scape.

\section{Results}

\subsection{Summary of model parameters }

We begin the discussion of our results by presenting a summary of
the parameters we have used for, and obtained from, our modelling.
These are given in Table 1. There, the mark OM refers to {\it
Obtained from modelling} and ST or {\it see text} refers to parameters
discussed in more detail in the previous Section on phenomenology,
where references are also given. These parameters values or ranges
of values are fixed by observations. Finally, the mark A refers to
{\it assumed parameters}, in general within a range. Variations to the values
given in Table 1 are discussed below.

\begin{table*}
\begin{center}
\centering \caption{Measured, assumed, and derived values for
different physical quantities at the innermost  starburst region
of NGC~253 (IS), a cylindrical disk with height 70 pc, and its
surrounding disk (SD).}
\begin{tabular}{lllll }
\hline
Physical parameters & Symbol & Value & Units & Comment\\
\hline

Distance & $D$ & 2.5 & Mpc & ST
\\

Inclination & $i$ & 78 & degrees & ST \\

Infrared Luminosity of the innermost starburst (IS)
& $L_{\rm IR}$  &$2\times 10^{10}$
 & L$_\odot$ & ST\\


Radius of the IS &-- & 100 & pc  &
ST\\

Radius surrounding disk (SD) &-- & 1000 & pc  &
ST\\

Uniform density  of the IS & $n_{\rm IS}$ & $\sim600$ & cm$^{-3}$
&
ST\\

Uniform density  of the SD & $n_{\rm SD}$ & $\sim50$ & cm$^{-3}$ &
ST\\

Gas mass of the IS & $M_{\rm IS}$ & $\sim 3 \times 10^7$ & M$_\odot$ & ST\\

Gas mass of the SD & $M_{\rm SD}$ & $\sim 2.5 \times 10^8$ & M$_\odot$ & ST\\

Supernova explosion rate of the IS & ${\cal R}$ & $\sim 0.08$ & SN
yr$^{-1}$ &
ST\\

Supernova explosion rate of the SD & -- & $\sim 0.01$ & SN
yr$^{-1}$ &
ST\\

Typical supernova explosion energy & -- &
10$^{51}$ & erg & ST\\

SN energy transferred to cosmic rays & $\eta$ &
$\sim 10$ & \% & ST \\

Convective velocity & $V$ & 300 -- 600 & km s$^{-1}$ & ST \\

Dust emissivity index & $\sigma$ & 1.5 & -- & OM \\

Dust temperature & $T_{\rm dust}$ & 50 & K & OM \\

Emission measure & EM & $5 \times 10^5$  & pc cm$^{-6}$ & OM \\

Ionized gas temperature & $ T$ &  $10^4$ & K & OM \\

Magnetic field of the IS & $B$ & 300 & $\mu$G & OM \\

Slope of primary injection spectrum & $p$ & 2.2--2.3 &-- & A\\

Maximum energy considered for primaries & -- & 100 & TeV & A
\\

Diffusion coefficient slope & $\mu$ & 0.5 &-- & A\\


Proton to electron primary ratio & $N_p/N_e$ & 50 & --  &
A\\

Diffusive timescale & $\tau_0$ & 1--10 & Myr & A\\

\hline

\end{tabular}
\label{model}
\end{center}
\end{table*}

\subsection{Steady proton and electron population}

\begin{figure*}[t]
\centering
\includegraphics[width=7cm,height=8cm]{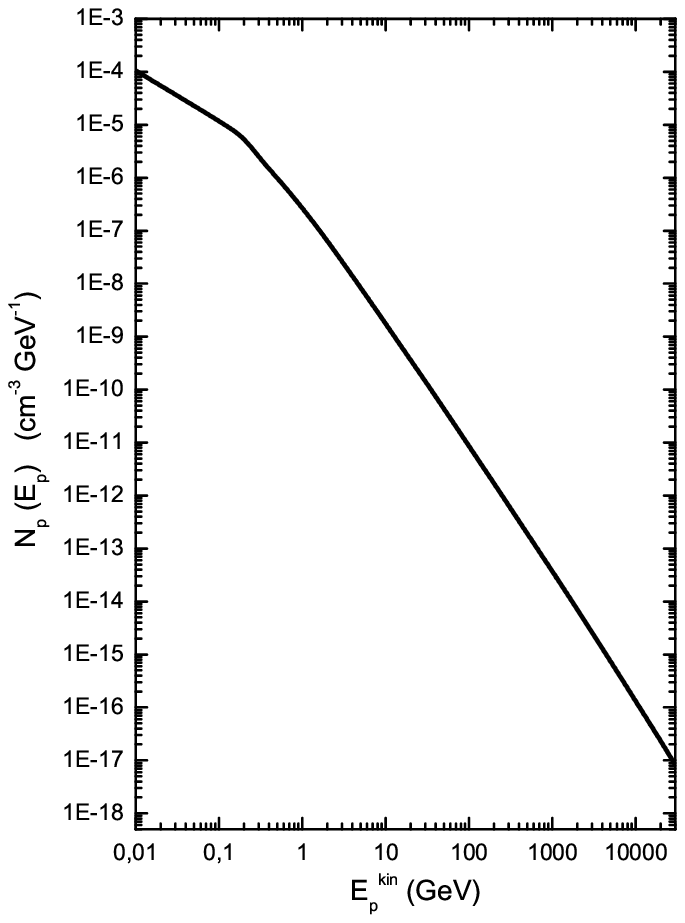}
\includegraphics[width=7cm,height=8cm]{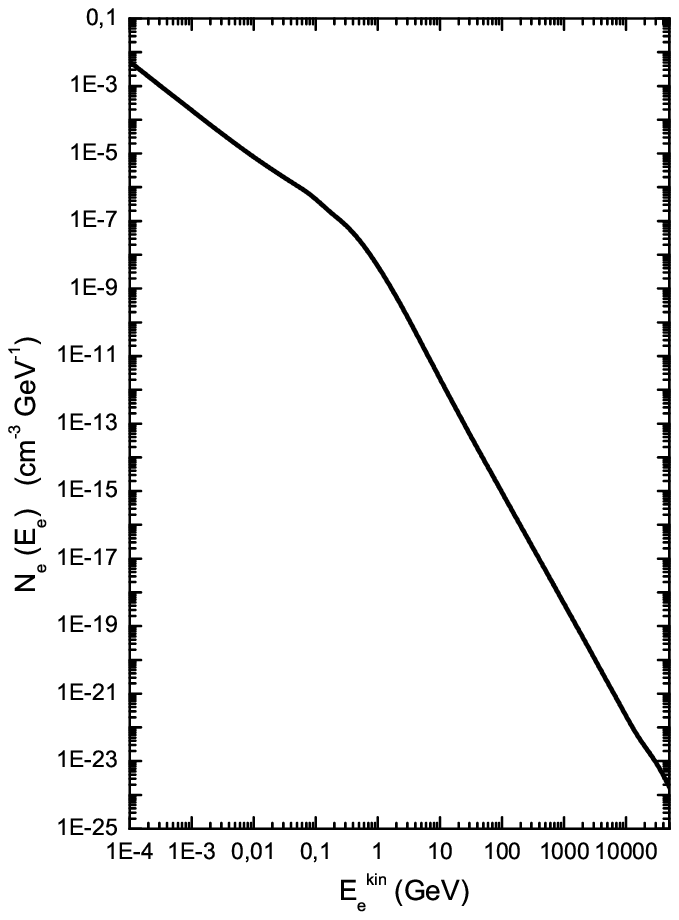}
 \caption{Steady proton (left panel) and electron (right
panel) distributions in the innermost region of NGC~253.}
\label{steady}
\end{figure*}

\begin{figure*}[t]
\centering
\includegraphics[width=7cm,height=8cm]{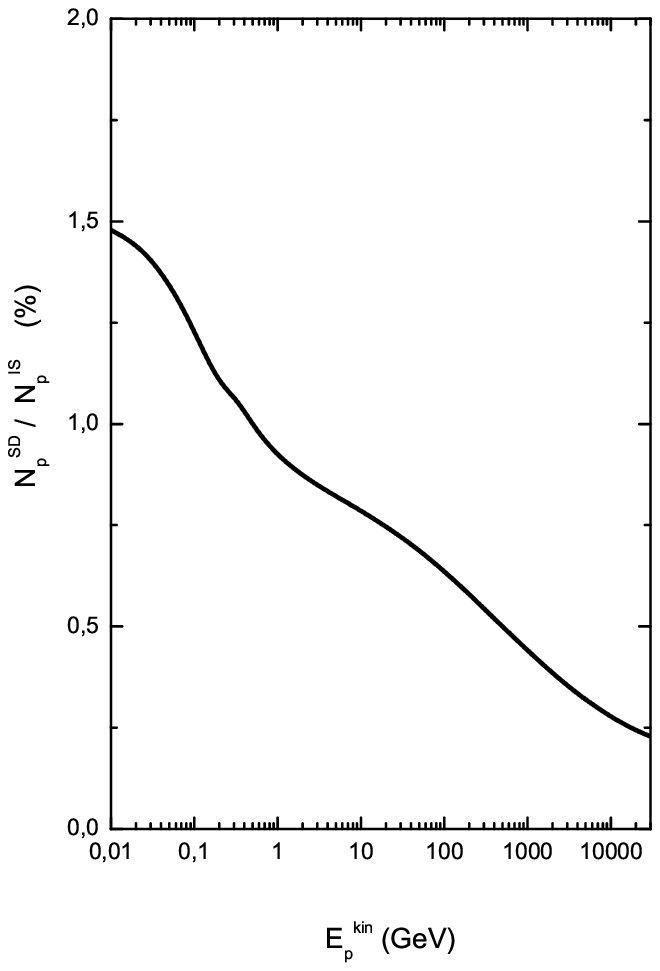}
\includegraphics[width=7cm,height=8cm]{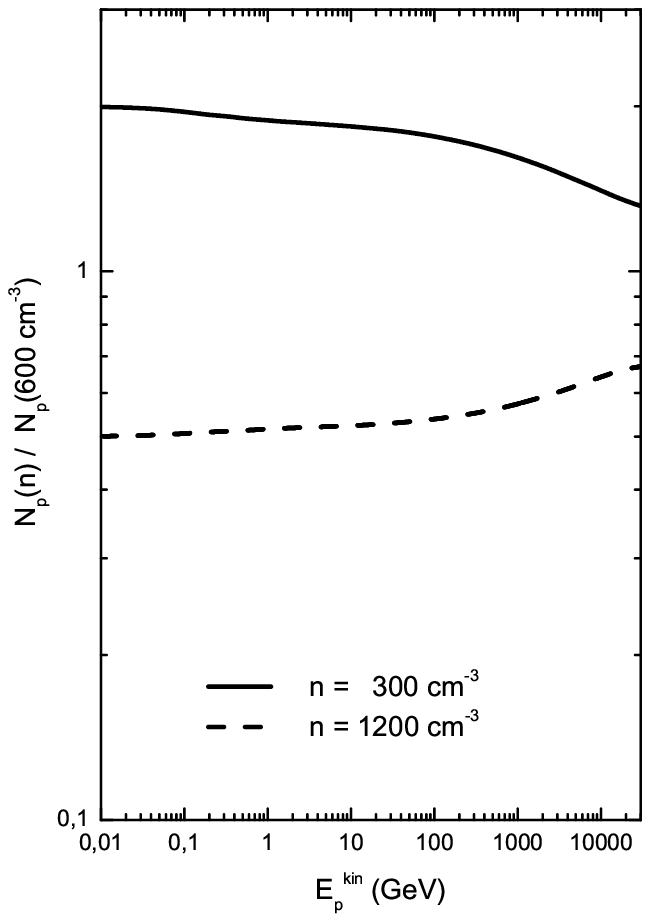}
 \caption{Left: Ratio of the steady proton population in
 the surrounding disk to that in the innermost starburst region.
 Right: Ratio between the steady proton distribution in the IS,
when the gas density is artificially enhanced
and diminished by a factor of 2.}
\label{steady2}
\end{figure*}

\begin{figure*}[t]
\centering
\includegraphics[width=7cm,height=8cm]{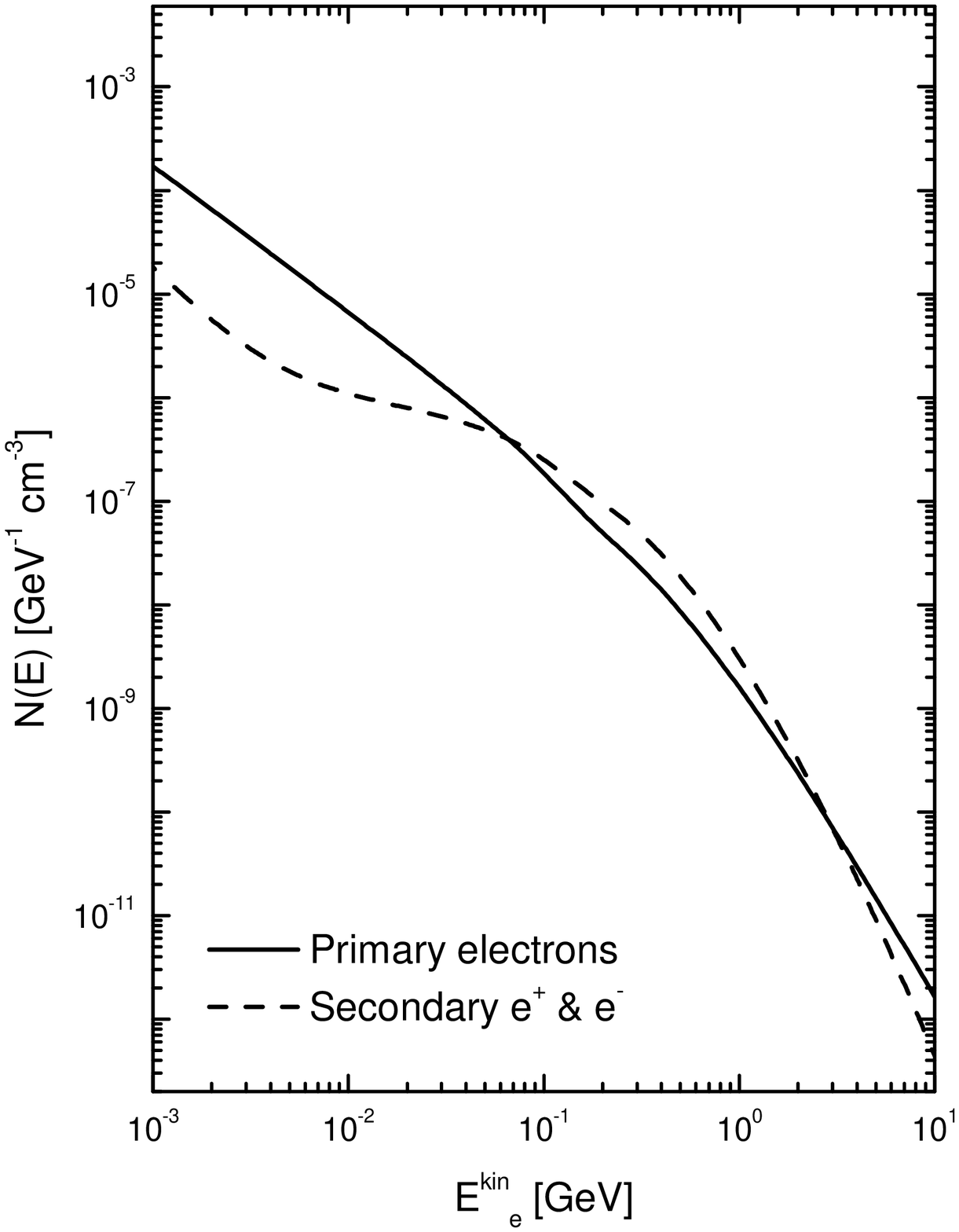}
 \caption{{Steady population of primary-only and secondary-only electrons. Only the region
 of the secondary dominance of the distribution is shown. }}
\label{primsec}
\end{figure*}

The numerical solution of the diffusion-loss equation for protons
and electrons, each subject to the losses previously described, is
shown in Figure \ref{steady}. We have adopted a diffusive
residence timescale of 10 Myr, a convective timescale of 1 Myr,
and a density of $\sim 600$ cm$^{-3}$. In the case of electrons,
the magnetic field with which synchrotron losses are computed in
Figure \ref{steady} is 300 $\mu$G. The latter is fixed below,
requiring that the steady electron population produces a flux
level of radio emission matching observations. An injection
electron spectrum is considered --in addition to the secondaries--
in generating the steady electron distribution. The primary
electron spectrum is assumed as that of the protons times a
scaling factor; the inverse of the ratio between the number of
protons and electrons, $N_p/N_e$ (e.g., Bell 1978). This ratio is
about 100 for the Galaxy, but could be smaller in star forming
regions, where there are multiple acceleration sites. For
instance, V\"olk et al. (1989) obtained $N_p/N_e \sim 30$ for M82.
$N_p/N_e=50$ is assumed for the central disk of NGC~253. From
about $E_e-m_e \sim 10^{-1}$ to $\sim 10$ GeV, the secondary
population of electrons (slightly) dominates, in any case. {This
is shown in Figure \ref{primsec}.} It is in this region of
energies where most of the synchrotron radio emission is
generated, and thus the ability of producing a high energy model
compatible with radio observations is a cross check for the
primary proton distribution.

Figure \ref{steady2} shows the ratio of the steady proton
population in the SD to that in the IS. Because the SD is subject
to a smaller supernova explosion rate, it has an smaller number of
protons in its steady distribution, at all energies, of the order
of 1\% of that of the IS. Then, it will play a subdominant role in
the generation of $\gamma$-ray emission, as we show below. In the
right panel of Figure \ref{steady2}, and for further discussion in
the following Sections, we present the ratio between the steady
proton distribution in the IS, when the gas density is
artificially enhanced and diminished by a factor of 2 from the
assumed value of 600 cm$^{-3}$.

\subsection{IR and radio emission}

\begin{figure*}[t]
\centering
\includegraphics[width=7cm,height=8cm]{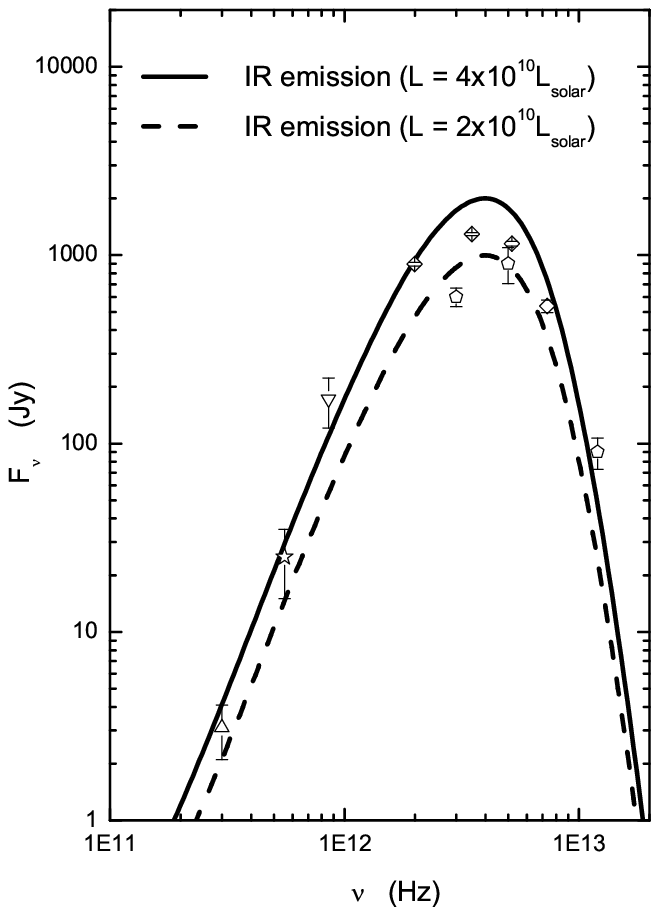}
\includegraphics[width=7cm,height=8cm]{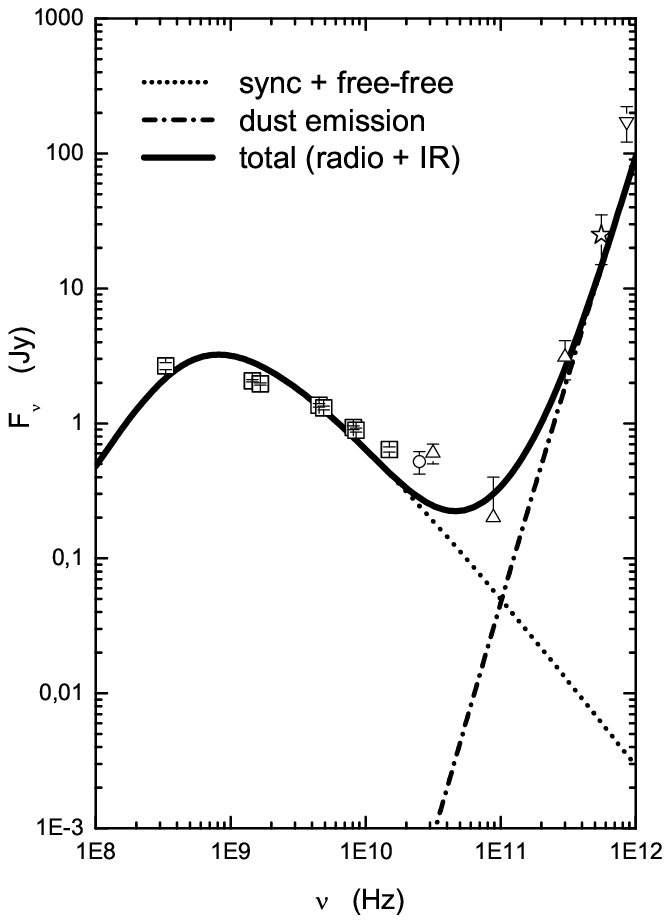}
\caption{Left: IR flux from NGC~253 assuming a
dilute blackbody with temperature $T_{\rm dust }= 50$ K and
different total luminosities. Right: Multifrequency spectrum of
NGC~253 from radio to IR, with the result of our modelling. The
experimental data points correspond to: pentagons, Melo et al. (2002);
diamonds, Telesco et al. (1980); down-facing triangles, Rieke et al. (1973);
stars, Hildebrand et al. (1977); up-facing triangles, Elias et al. (1978);
circles, Ott et al. (2005); squares, Carilli (1996).} \label{IR-radio}
\end{figure*}

The continuum emission from NGC~253, at wavelengths between $\sim
1$ cm and $\sim 10$ microns, was measured by several authors,
e.g., see Figure \ref{IR-radio} and Section 3. These observations
did not in general distinguish, due to angular resolution, only
the emission coming from the innermost starburst region. Instead,
they also contain a contribution coming from photons produced in
the surrounding disk and farther away from the nucleus. The IR
continuum emission is mainly produced thermally, by dust, and thus
it could be modelled with a spectrum having a dilute blackbody
(graybody) emissivity law, proportional to
$\nu^{\sigma}B(\epsilon, T)$, where $B$ is the Planck function.
Figure \ref{IR-radio} shows the result of this modelling and its
agreement with observational data when the dust emissivity index
$\sigma=1.5$ and the dust temperature $T_{\rm dust}= 50 K$.
Different total luminosities, the normalization of the dust
emission (see the appendix of Torres 2004 for details), are shown
in the Figure to give an idea of the contribution of the innermost
region with respect to that of the rest of the galaxy. According
to Melo et al. (2002), about half of the total IR luminosity is
produced in the IS, what is consistent with the data points being
intermediate between the curves with $L_{\rm IR}$ 2$\times
10^{10}$ and 4$\times 10^{10}$~L$_\odot$, since the latter were
obtained with beamsizes of about 20--50 arcsec ($\sim$ 240--600 pc
at the NGC~253 distance).

The influence of the magnetic field upon the steady state electron
distribution is twofold. On one hand, the greater the field, the
larger the synchrotron losses --what is particularly visible at
high energies, where synchrotron losses play a relevant role. On
the other, the larger the field the smaller the steady
distribution. These effects evidently compete between each other
in determining the final radio flux. The magnetic field is
required to be such that the radio emission generated by the
steady electron distribution is in agreement with the
observational radio data. This is achieved by iterating the
feedback between the choice of magnetic field, the determination
of the steady distribution, and the computation of radio flux,
additionally taking into account free-free emission and absorption
processes. Whereas free-free emission is subdominant when compared
with the synchrotron flux density, free-free absorption plays a
key role at low frequencies, determining the opacity. We
have found a reasonable agreement with all observational data for
a magnetic field in the innermost region of 300 $\mu$G, an ionized
gas temperature of about 10$^4$ K, and an emission measure of $5
\times 10^5$ pc cm$^{-6}$, the latter two are in separate
agreement with the free-free modelling of the opacity of
particular radio sources, as discussed in Section 3. The value of
magnetic field we have found for the IS is very similar to that
found for the disk of Arp 220 (Torres 2004) and compatible with
measurements in molecular clouds (Crutcher 1988, 1994, 1999).

\subsection{$\gamma$-ray emission}

\begin{figure*}[t]
\centering
\includegraphics[width=7cm,height=8cm]{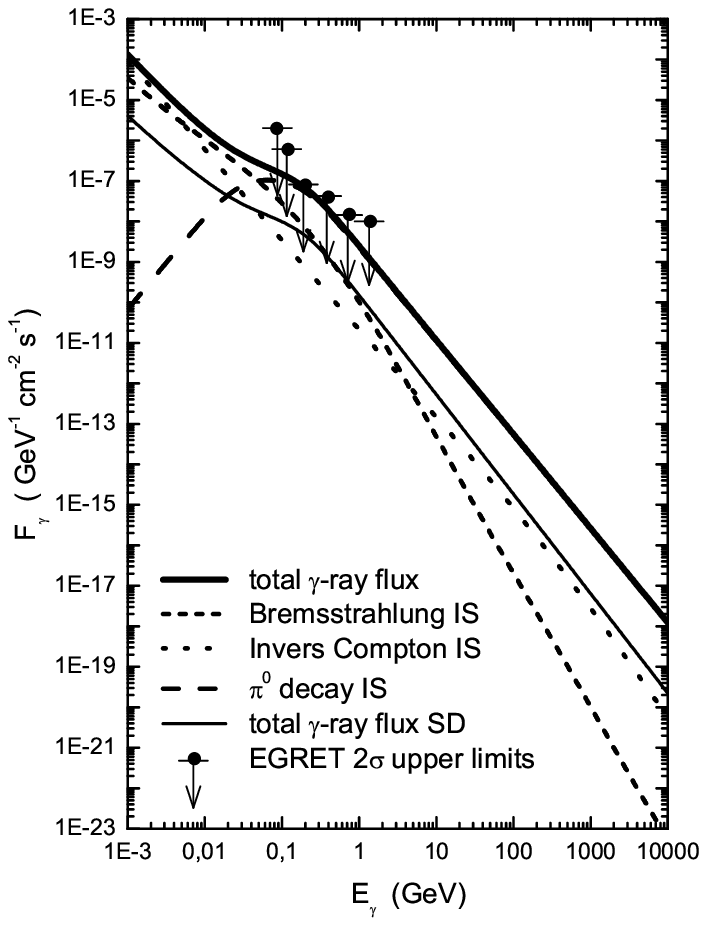}
\includegraphics[width=7cm,height=8cm]{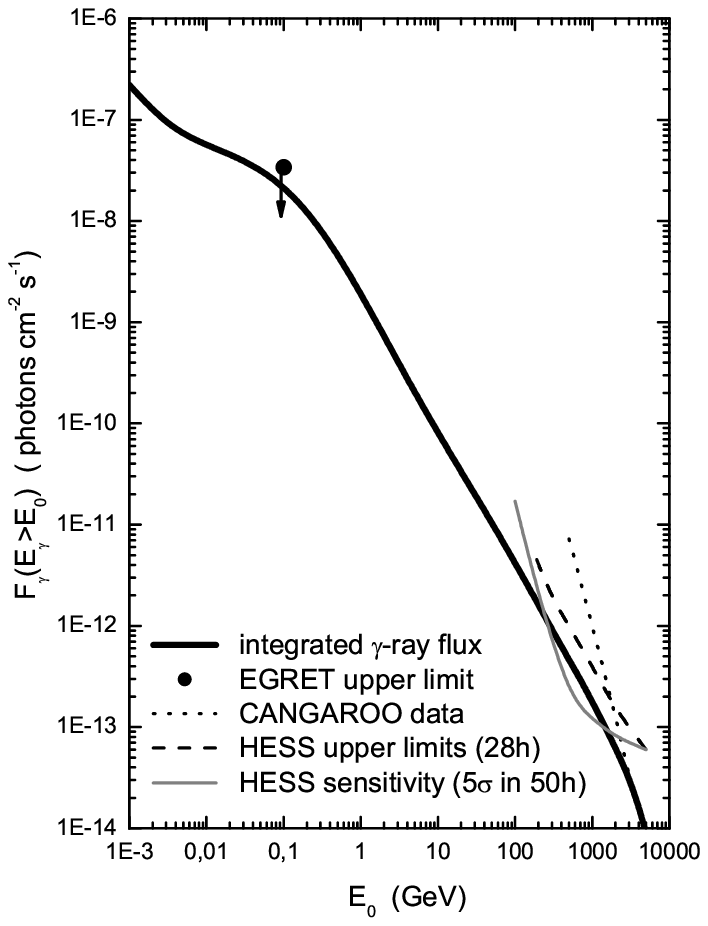}
\caption{Left: Differential $\gamma$-ray fluxes from the central
region of NGC~253. Total contribution of the surrounding disk is
separately shown, as are the EGRET upper limits. In the case of
the IS, we separately show the relative contributions of
bremsstrahlung, inverse Compton, and neutral pion decay to the
$\gamma$-ray flux. Right: Integral $\gamma$-ray fluxes. The EGRET
upper limit (for energies above 100 MeV), the CANGAROO integral
flux as estimated from their fit, and the HESS sensitivity (for a
5$\sigma$ detection in 50 hours) are given. Absorption effects are
already taken into account. Also shown is the recently released
HESS upper limit curve on NGC 253. } \label{g-emis}
\end{figure*}

In the left panel of Figure \ref{g-emis}, bremsstrahlung, inverse
Compton, and pion decay $\gamma$-ray fluxes of the IS are shown
together with the total contribution of the SD and the total
differential flux of the whole system. These results are
obtained with the model just shown to be in agreement with radio
and IR observations. As mentioned before, the contribution of the SD, even when having
a factor of $\sim 8$ more mass than the IS, is subdominant.
The reason for this needs to be found
in that this region is much less active (Ulvestad 2000).

Our predictions, while complying with EGRET upper limits, are
barely below them. If this model is correct, NGC~253 is bound
to be a bright $\gamma$-ray source for GLAST.

The integral fluxes are shown in the right panel of Figure
\ref{g-emis}. Our model complies again with the integral EGRET
upper limit for photons above 100 MeV, and predicts that, given
enough observation time, NGC~253 is also to appear as a point-like
source in an instrument like HESS. Note, however, that quite a
long exposure may be needed to detect the galaxy, and also, that
our fluxes are only a few percent of those reported by the
CANGAROO collaboration.

An additional source of TeV photons not considered here is the
hadronic production in the winds of massive stars (Romero \&
Torres 2003). However, a strong star forming region such as the
nucleus of NGC~253 is bound to possess much more free gas than
that contained within the winds of massive stars, which albeit
numerous, have mass loss rates typically in the range of
10$^{-6}$--10$^{-7}$ M$_\odot$.\footnote{In Romero \& Torres
(2003), higher mass loss rates up to $10^{-5}$ M$_\odot$, i.e.,
grammages between 50 and 150 g cm$^{-2}$ were allowed. These
values, although have been found in perhaps one or two Galactic
early O stars, are uncommon. Since the size of the base of the
wind for each star, the grammage, and the ambient enhancement of
cosmic rays were independently allowed to take values within their
assumed ranges in the Monte Carlo simulation of Romero \& Torres
(2003), the stars with the most favorable parameters would
dominate the sum, overestimating the relative importance of their
fluxes.}

\subsection{Opacities to $\gamma$-ray escape}

\begin{figure*}[t]
\centering
\includegraphics[width=7cm,height=8cm]{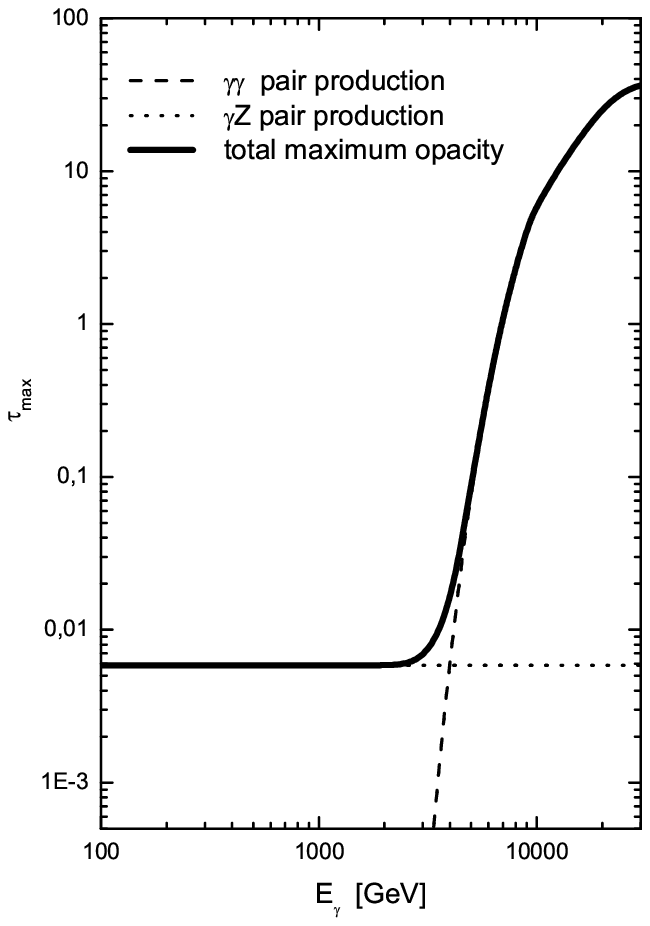}
\includegraphics[width=7cm,height=8cm]{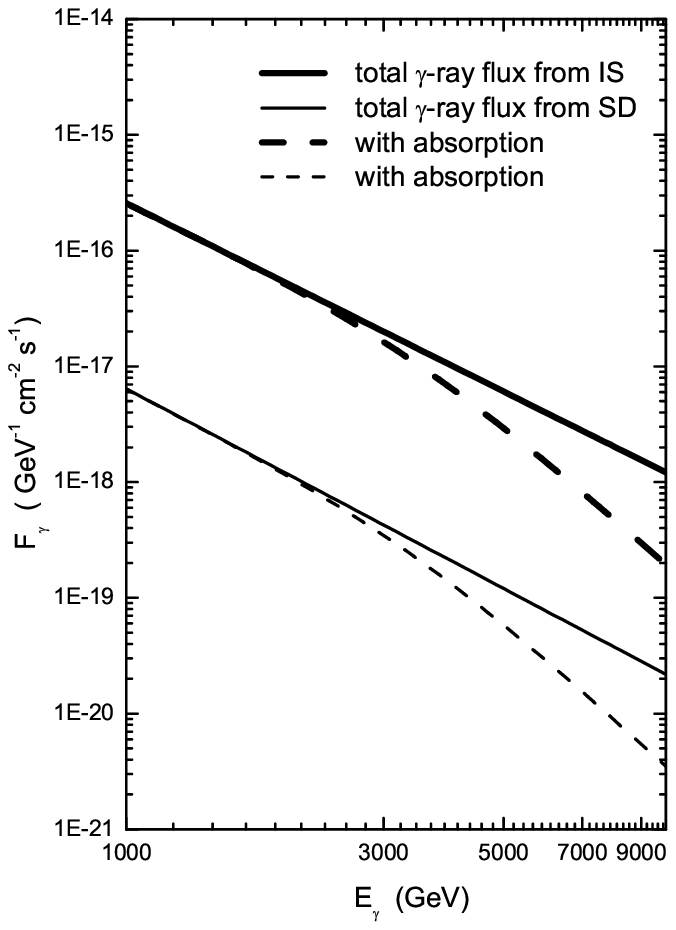}
\caption{Left: Opacities to $\gamma$-ray scape as a function of
energy. The highest energy is dominated by $\gamma\gamma$
processes, whereas $\gamma Z$ dominates the opacity at low
energies. Significant $\tau_{\rm max}$ are only encountered above
1 TeV. Right: Modification of the $\gamma$-ray spectrum introduced
by the opacity to $\gamma$-ray escape. } \label{opa}
\end{figure*}

Two sources of opacities are considered: pair production from
the $\gamma$-rays interaction with the photon field or with the
charged nucleus present in the medium. Compton scattering and
attenuation in the magnetic field by one-photon pair production
are negligible.

The opacity to $\gamma\gamma$ pair production
with the photon field, which, at the same time, is target for
inverse Compton processes, can be computed as $
\tau(R_{c},E_\gamma)^{\gamma\gamma}=\int \int_{R_c}^\infty
n(\epsilon) \sigma_{e^-e^+}(\epsilon,E_\gamma)^{\gamma\gamma}  dr
\, d\epsilon < \tau(E_\gamma)_{\rm max}^{\gamma\gamma} < (h/ \cos
(i) ) \int_0^\infty n(\epsilon) \sigma_{e^-e^+}(\epsilon,E_\gamma)
d\epsilon \,  $ where $\epsilon$ is the energy of the target
photons, $E_\gamma$ is the energy of the $\gamma$-ray in
consideration, $R_c$ is the place where the $\gamma$-ray photon
was created within the system, and $
\sigma_{e^-e^+}(\epsilon,E_\gamma)^{\gamma\gamma} =
({3\sigma_T}/{16}) (1-\beta^2) (2\beta
(\beta^2-2)+(3-\beta^4)\ln((1+\beta)/(1-\beta))), $ with
$\beta=(1-(m c^2)^2/(\epsilon \, E_\gamma))^{1/2}$ and $\sigma_T$
being the Thomson cross section, is the cross section for
$\gamma\gamma$ pair production (e.g. Cox 1999, p.214). Note that
the lower limit of the integral on $\epsilon$ in the expression
for the opacity is determined from the condition that the center
of mass energy of the two colliding photons should be such that
$\beta >0$. The fact that the dust within the starburst reprocesses the UV
star radiation to the less energetic infrared photons implies that
the opacities to $\gamma\gamma$ process is significant only at the
highest energies. No source of this kind of opacity is assumed outside
the system under consideration, since the nearness of NGC~253 makes negligible
the
opacity generated by cosmological fields.

The cross section for pair production from the interaction of a
$\gamma$-ray photon in the presence of a nucleus of charge $Z$
in the completely screened regime ($E_\gamma / mc^2 \gg
1/(\alpha Z)$) is independent of energy, and is given by (e.g. Cox
1999, p.213) $ \sigma_{e^-e^+}^{\gamma Z} = (3 \alpha Z^2 \sigma_T
/ 2 \pi ) (7/9 \ln (183/Z^{1/3}) - 1/54)$. At lower energies the
relevant cross section is that of the no-screening case, which has
a logarithmic dependency on energy, $ \sigma_{e^-e^+}^{\gamma Z} =
(3 \alpha Z^2 \sigma_T / 2 \pi ) (7/9 \ln (2E_\gamma/mc^2) -
109/54)$, and matches the complete screening cross section at
around 0.5 GeV. Both of these expression are used to compute the
opacity, depending on $E_\gamma$.

In the left panel of Figure \ref{opa} we show the different
contributions to the opacity. The equation of radiation transport
appropriate for a disk is used to compute the predicted
$\gamma$-ray flux taking into account all absorption processes
(see Appendix of Torres 2004 for details). The right panel of
Figure \ref{opa} shows the effect of the opacity on the integral
$\gamma$-ray fluxes, only evident above 3 TeV.

\subsection{Exploring the parameter space and degeneracies}

\begin{figure*}[t]
\centering
\includegraphics[width=7cm,height=8cm]{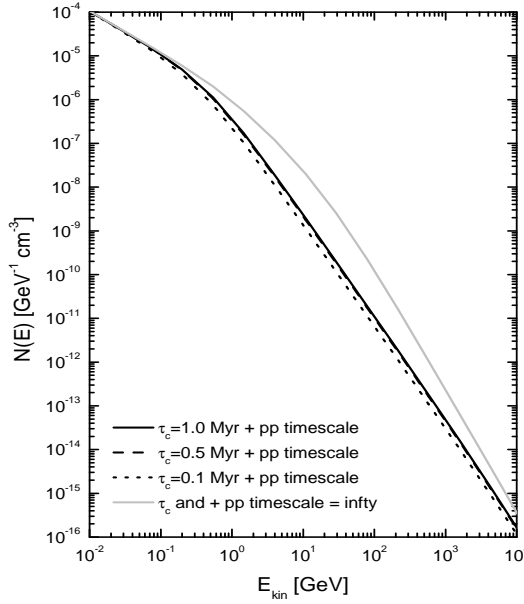}
\caption{Proton distribution when different convective and the pp
timescales are taken into account as compared with a (arbitrary)
solution when $\tau(E)=\tau_D$, i.e., diffusion only. Clearly,
convection plus pp timescales dominates the spectrum. }
\label{timescales}
\end{figure*}

As it is summarized in Table 1, most of the model parameters are
well fixed from observations. There are however a couple of
assumptions which, whereas being not well bounded from
experiments, may produce slight degeneracies. Consider for
instance the proton injection slope $p$ and the { diffusive} scale
$\tau_0$. For the former we have assumed $p=2.3$, which is in
agreement with the recent results from HESS regarding $\gamma$-ray
observations at TeV energies of supernova remnants and
unidentified extended sources. However, it would be certainly
within what one would expect from proton acceleration in supernova
remnants, and also within experimental uncertainty, if a better
description for the average proton injection slope in NGC~253 is
2.2 instead of 2.3. Table \ref{deg} shows the influence of this
kind of choice on our final results. A harder slope slightly
increases the integral flux. Similarly, the { diffusive} timescale
is not well determined, and it may be arguable perhaps within one
order of magnitude. Table \ref{deg} also shows the influence of
this choice. Ultimately, high energy $\gamma$-ray observations
(from GeV to TeV) are the ones to impose constraints on these
values. { In any case, we remark that pp interaction and
convection timescales are much shorter ($< 1$ Myr) and thus
dominate the form of $N(E)$. To show this in greater detail we
show in Figure \ref{timescales} the result for the proton
distribution when different convective and the pp timescales are
taken into account as compared with the solution when
$\tau(E)=\tau_D$, i.e., diffusion only. Clearly, convection plus
pp timescales dominates the spectrum.}

Regarding degeneracies, both the proton slope and the confinement
timescales, however, cannot be much different from what we have
assumed. If the former were to differ significantly, it would be
impossible to reproduce the radio data, which is the result of the
synchrotron emission of the secondary electrons. Changes in the
number of protons in the IS would imply a change in the magnetic
field to reproduce radio observations, what clearly cannot be
pushed much either.

In a less impacting way, varying the value of $N_p/N_e$ can also
vary the results. This variation would be slight because of the
influence of the more numerous secondary electrons in the
energetic region of interest for radio emission. On the same
track, varying the diffusion coefficient $\mu$ does not import
substantial changes. And if the maximum proton energy were to
differ from the value of 100 TeV we have assumed (what we do not
expect it to happen in a significant way, since we do now
observationally know that supernova remnants are sources of $\sim
10$ TeV photons), the end of the spectrum would slightly shift
accordingly.

\begin{table*}
\begin{center}
\centering \caption{Exploring the parameter space for $p$ and
$\tau_0$. The results of our adopted model are given in the first
column. These results already take into account the opacity to
photon escape.}
\begin{tabular}{lllllll }
\hline
$F(E > E_{0})$ & $ p=2.3$        & $ p=2.3$        & $ p=2.3$        & $ p=2.2$          & $ p=2.2$       & $ p=2.2$   \\
                & $\tau_0=10$ Myr & $\tau_0=5$ Myr & $\tau_0=1$ Myr & $\tau_0=10$ Myr   & $\tau_0=5$ Myr & $\tau_0=1$ Myr \\
 \hline

$E_{0}=100$ MeV  &  2.32E-8   & 2.36E-8  & 2.21E-8  &  2.95E-8   &    2.97E-8  &   2.75E-8\\
$E_{0}=200$ GeV  &  1.60E-12  & 1.23E-12 & 4.76E-13 &  4.04E-12  &    3.08E-12 &   1.15E-12\\
$E_{0}=600$ GeV  &  3.61E-13  & 2.67E-13 & 8.98E-14 &  1.00E-12  &    7.34E-13 &   2.40E-13\\
$E_{0}=1$ TeV    &  1.78E-13  & 1.29E-13 & 4.10E-14 &  5.16E-13  &    3.70E-13 &   1.14E-13\\
$E_{0}=2$ TeV    &  6.29E-14  & 4.46E-14 & 1.31E-14 &  1.92E-13  &    1.35E-13 &   3.87E-14\\

\hline

\end{tabular}
\label{deg}
\end{center}
\end{table*}

Even within an artificially enlarged
uncertainty of the gas density, the results will not be modified much:
if for any reason the average particle density were
to be a factor of 2 smaller or larger, the $\gamma$-ray integral
flux variations would be within  4\% for energies above 100 MeV,
and within  25\% for energies above 200 GeV. Table \ref{density}
shows these results by presenting the integral fluxes above a given
threshold if the assumed density of 600 cm$^{-3}$ is doubled or halved.
As can be seen in the right panel of Figure
\ref{steady2}, if the density is larger (smaller) by a factor $\sim 2$,
the resultant steady proton distribution from the same
proton injection population is smaller (bigger) by a similar factor over a
wide range of proton energies. As $\gamma$-ray
emissivities are proportional to both the medium density and the
number of steady protons, the variations in $\gamma$-ray fluxes
are greatly compensated.

\begin{table*}
\begin{center}
\centering \caption{The effect of the medium gas density on the
$\gamma$-ray integral fluxes. Results provided are in units of
photons cm$^{-2}$ s$^{-1}$, and already take into account
the opacity to photon escape.}
\begin{tabular}{llll }
\hline
$F(E > E_{0})$   &   $n$=300 cm$^{-3}$  &  $ n$=600 cm$^{-3}$  &  $ n$=1200 cm$^{-3}$   \\
\hline
$E_{0}=100$ Mev  &  2.22E-8   &  2.32E-8   &  2.37E-8    \\
$E_{0}=200$ Gev  &  1.20E-12  &  1.60E-12  &  1.95E-12   \\
$E_{0}=600$ Gev  &  2.60E-13  &  3.61E-13  &  4.52E-13   \\
$E_{0}=1$ Tev    &  1.26E-13  &  1.78E-13  &  2.26E-13   \\
$E_{0}=2$ Tev    &  4.36E-14  &  6.29E-14  &  8.09E-14   \\
\hline
\end{tabular}
\label{density}
\end{center}
\end{table*}

\subsection{Energetics and cosmic ray enhancement}

\begin{figure*}[t]
\centering
\includegraphics[width=6cm,height=7cm]{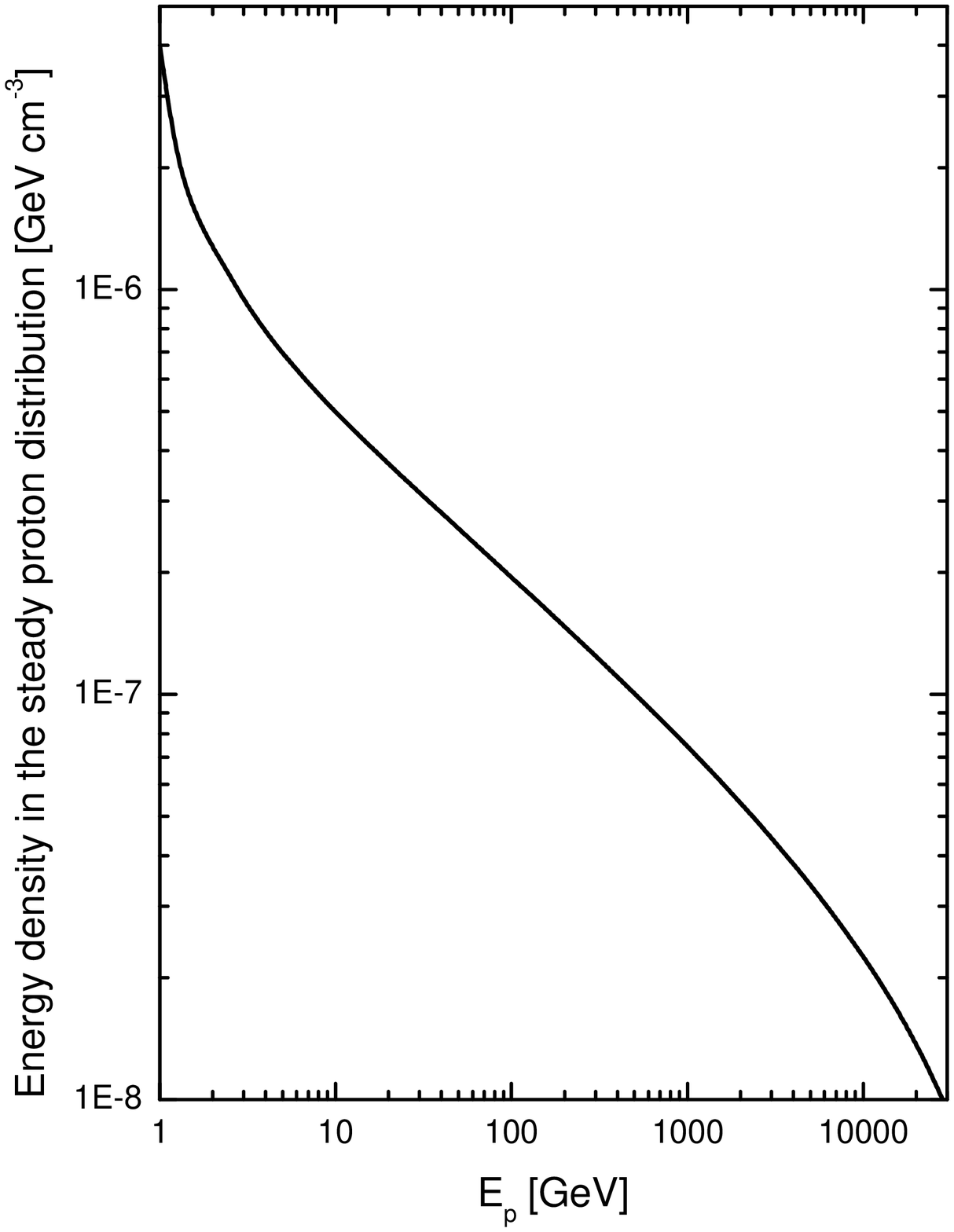}
\includegraphics[width=6cm,height=7cm]{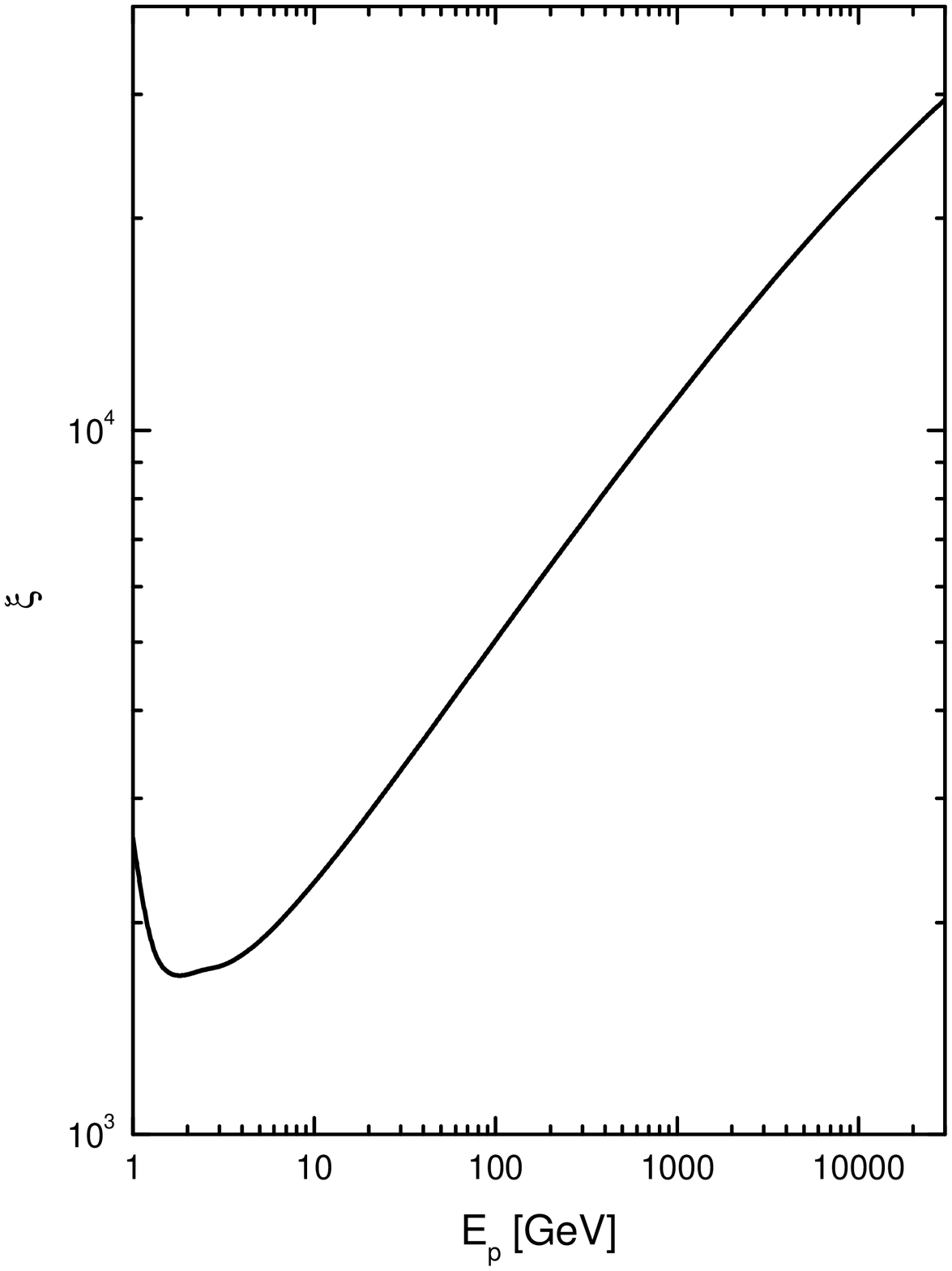}
\caption{Left: Energy density contained in the steady population
of protons above the energy given by the $x$-axis. Right: Cosmic
ray enhancement factor obtained from the steady spectrum
distribution in the innermost starburst nucleus of NGC~253.}
\label{varsigma}
\end{figure*}

The left panel of Figure \ref{varsigma} presents the energy
density contained in the steady proton population above a certain
energy, i.e., based on Figure \ref{steady}, the curve shows the
integral $\int_E N_{p}(E_p) \, E_p \, dE_p $. The total energy
density contained by the steady population of cosmic rays above 1
GeV is about $10^{-3}$ of the power emitted by all supernova
explosions in the last 5 million years. The energy density
contained in the steady electron population is orders of magnitude
less important.

The cosmic ray enhancement is a useful parameter in estimations of
$\gamma$-ray luminosities in different scenarios. It is defined as
the increase in the cosmic ray energy density with respect to the
local value, $ \omega_{\rm CR,\odot}(E)=\int_E N_{p\,\oplus} (E_p)
E_p dE_p $, where $N_{p\,\oplus}$ is the  local cosmic ray
distribution obtained from the measured cosmic ray flux that we
quote in the Appendix. The enhancement factor $\varsigma$ is then
a function of energy given by $\varsigma(E)=(\int_E N_{p}(E_p) \,
E_p \, dE_p ) /\omega_{\rm CR,\odot}(E)$. Values of enhancement
for NGC~253 were proposed  $ \varsigma \leq 3000$ for energies
above 1 GeV (e.g., Suchkov et al. 1993), and we can actually
verify this in our model. The right panel of Figure \ref{varsigma}
presents the enhancement factor as a function of proton energy.
The larger the energy, the larger the enhancement, due to the
steep decline ($\propto E^{-2.75}$) of the local cosmic ray
spectrum.

\section{HESS observations}

\begin{figure*}[t]
\centering
\includegraphics[width=7cm,height=8cm]{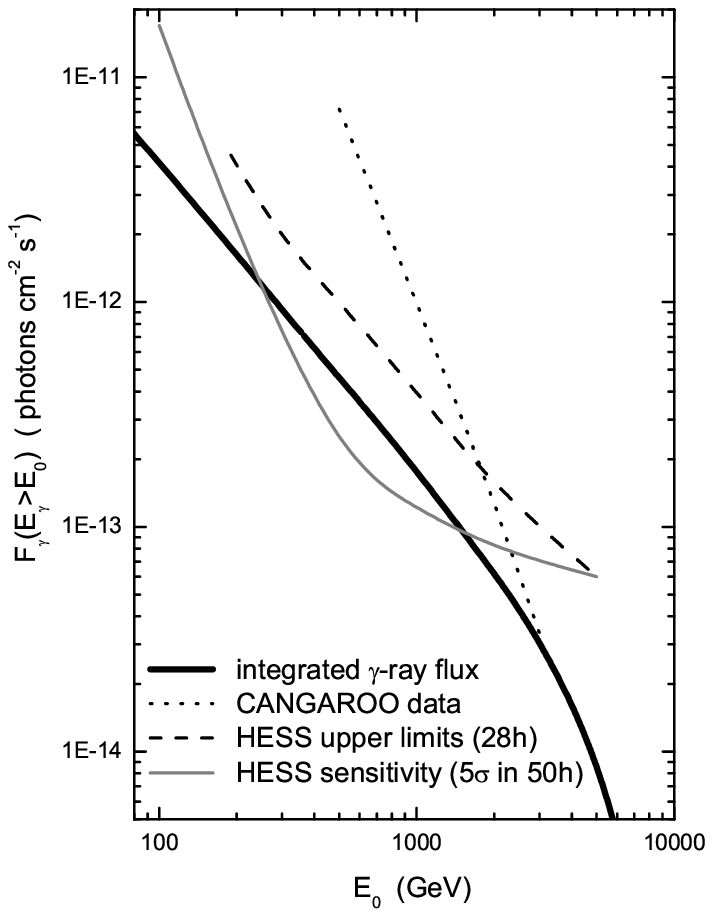}
\caption{{ Integral $\gamma$-ray fluxes zoomed in the region above
100 GeV, together with the (2 telescopes) HESS upper limits and
the (4 telescopes) HESS sensitivity.}} \label{g-emis2}
\end{figure*}

{The HESS array has just released (Aharonian et al. 2005c) their
results for NGC 253. These are based on data taken during the
construction of the array with 2 and 3 telescopes operating. The
total observation time was 28 hs, with a mean zenith angle of
about 14 degrees. Only events where at least two telescopes were
triggered were used, to enable stereoscopic reconstruction. The
energy threshold for this dataset was 190 GeV. Upper limits from
H.E.S.S. on the integral flux of $\gamma$-rays from NGC 253 (99 \%
confidence level) are shown, together with our predictions, in the
right panel of Figure \ref{g-emis}, and zoomed in the region above
100 GeV in Figure \ref{g-emis2}. As an example, above 300 GeV, the
upper limit is $1.9 \times 10^{-12}$ photons cm$^{-2}$ s$^{-1}$.
It can be seen that our predictions are below these upper limits
at all energies but still above HESS sensitivity for reasonable
observation times.}

\section{Concluding remarks}

We have presented a  multifrequency model of the central region of
NGC~253. Following recent observations, we have modelled an
innermost starburst with a radius of 100 pc and a supernova
explosion rate of 0.08 yr$^{-1}$, and a surrounding disk up to a 1
kpc in radius with an explosion rate about tenfold smaller. As a
result of our modelling we have found that a magnetic field of 300
$\mu$G for the innermost region is consistent with high resolution
radio observations, with the radiation at 1 GHz being mostly
produced by secondary electrons of cosmic ray interactions. The
magnetic field found for the innermost part of NGC~253 is typical
of dense molecular clouds in our Galaxy, and is close to the  (270
$\mu$G) value proposed by Weaver et al. (2002) using the
equipartition argument. We have estimated free-free emission and
absorption, and considered opacities to the $\gamma$-ray escape.
 The hard X-ray emission from IC and bremsstrahlung processes
produced in this model is below observational constraints, e.g.,
by OSSE, in agreement with previous estimations of bremsstrahlung
diffuse emission (Bhattacharya et al. 1994). This is consistent
with measurements in the center of the Galaxy, where INTEGRAL have
shown that hard X-ray emission is not diffusively, but produced by
point like sources (Lebrun et al. 2004).

The flux predicted is based on a set of a few well founded
assumptions, mainly a) that supernova remnants accelerate most of
the cosmic rays in the central region of NGC~253, and b) that they
interact with the present gas, whose amount has been measured
using a variety of techniques. The low opacity to $\gamma$-ray
escape secure that basically all $\gamma$-rays produced in the
direction towards Earth reach us. Observational constraints
establishes the values of the supernova explosion rate and gas
content (see Section 2 for references).

The ease of all the assumptions made in our model, its concurrence
with all observational constraints, and the unavoidability of the
processes analyzed, lead us to conclude that 1) GLAST will detect
NGC~253, being our predicted luminosity ($2.3 \times 10^{-8}$
photons cm$^{-2}$ s$^{-1}$ above 100 MeV) well above its 1 yr all
sky survey sensitivity (GLAST Science Requirements Doc. 2003);  2)
that our predicted TeV fluxes are about one order of magnitude
smaller than what was claimed by CANGAROO, and thus, that perhaps
in this case, a similar problem to that found in other sources
affected their data taking or analysis, and 3) that HESS could
detect the galaxy as a point like source provided it is observed
long enough with the full array ($\gtrsim 50$ hours, for a
detection between 300 and 1000 GeV.) \footnote{HESS site latitude
provides that NGC~253 can be observed very close to the zenith
(the minimum zenith angle for NGC~253 from HESS site is 2
degrees). As a consequence, HESS observations of NGC~253 can be
done with the minimum energy threshold of the experiment. The
MAGIC Telescope, although being at a northern hemisphere site, is
also able to observe NGC~253 at a larger zenith angle, about 53
degrees.}
We finally note that this model predicts a steady $\gamma$-ray
source, so that a posteriori variability estimators (e.g., Torres
et al. 2001) can be checked for consistency.

\section*{Appendix: Parameterizations
of proton-proton cross sections for neutral pion decay}

The $\pi^0$ emissivity resulting from an isotropic intensity of
protons, $J_p(E_p)$, interacting with fixed target nuclei with
number density $n$, through the reaction $p+p \rightarrow
p+p+\pi^0 \rightarrow p+p+2\gamma $, is given by (e.g., Stecker
1971) \be Q_{\pi^0}(E_{\pi^0}) = 4 \pi n
\int_{E_{th}(E_{\pi^0})}^{E_p^{max}} dE_p\, J_p(E_p)
{d\sigma(E_{\pi^0}, E_p)\over dE_{\pi^0}}\;, \label{1} \ee where
$E_p^{max}$ is the maximum energy of protons in the system, and
$E_{th}(E_{\pi^0})$ is the minimum proton energy required to
produce a pion with total energy $E_{\pi^0}$, and is determined
through kinematical considerations. It is obtained using the
invariant, $\sqrt{s}= \left( 2m_p (E_p+m_p)\right)^{1/2} = (M_X^2
+ {E_\pi^*}^2 - m_\pi^2)^{1/2} +{E_\pi ^*}^2$, where $s$ is the
square of the total energy in the center-of-mass system, $M_X$
depends on the reaction channel and represents the invariant mass
of the system consisting of all particles except the pion,
${E_\pi^*}$ is the CMS energy of the produced meson (that is
connected with the laboratory  system energy via a Lorentz
transformation, see Appendices of Moskalenko \& Strong 1998 and
Blattnig et al. 2000b). ${E_\pi^*}=(s-M_x^2+m_\pi^2)/(2\sqrt{s})$,
so that for a given value of $s$, ${E_\pi^*}$ will be maximum when
$M_x$ takes its minimum value. For the case of neutral pion
production $M_x=2\,m_p$. The laboratory system pion energy,
obtained from ${E_\pi^*}$, can be put as a function of $s$.
Inverting this relation, thus obtaining $s=s(E_\pi)$, and use of
$s=s(E_\pi)=2m_p(E_p+m_p)$ allow for the minimum proton energy to
be derived. Finally, ${d\sigma(E_{\pi^0}, E_p) / dE_{\pi^0}}$ is
the differential cross section for the production of a pion with
energy $E_{\pi^0}$, in the lab frame, due to a collision of a
proton of energy $E_p$ with a hydrogen atom at rest.

The $\gamma$-ray emissivity is obtained from the neutral pion
emissivity $Q_{\pi^0}$ as \be \label{pion-prog}
{Q_\gamma(E_\gamma)}_{\pi} = 2 \int_{E_{\pi^0}^{min} (E_\gamma)}
^{E_{\pi^0}^{max} (E_p^{max})} dE_{\pi^0} {Q_{\pi^0}(E_{\pi^0})
\over (E_{\pi^0}^2 - m_{\pi^0}^2 c^4)^{1/2}} \ee
where $E_{\pi^0}^{min} (E_\gamma) = E_\gamma + m_{\pi^0}^2 c^4 /
(4E_\gamma)$ is the minimum pion energy required to produce a
photon of energy $E_\gamma$ (e.g., Stecker 1971), and
$E_{\pi^0}^{max} (E_p^{max})$ is the maximum pion energy that the
population of protons can produce. It is obtained using the
invariant equation for the maximum proton energy resident in the
system.

Thus, an accurate knowledge of the differential cross section for
pion production becomes very important to estimate the
$\gamma$-ray emissivity. Note that ${d\sigma(E_{\pi^0},
E_p)/dE_{\pi^0}}$ can be thought as containing the inclusive total
inelastic cross section (i.e. the cross section multiplied by the
average pion multiplicity). This can be stated explicitly as done,
for instance, in Dermer's (1986a) equation 3.

\subsection*{$\delta$-function approximation}

In this formalism (Aharonian and Atoyan 2000), \ba
Q_{\pi^0}(E_{\pi^0}) = 4 \pi n \int_{E_{th}(E_{\pi^0})} dE_p\,
J_p(E_p) \, \delta(E_{\pi^0}-\kappa  E_{\rm kin})\,
\sigma(E_p)\;, \nonumber \\
=\frac{ 4 \pi n}{\kappa} J_p\left(m_p c^2+
\frac{E_{\pi^0}}{\kappa}\right) \sigma\left(m_p c^2+
\frac{E_{\pi^0}}{\kappa}\right) \label{delta} \ea where $\sigma$
is the total cross-section of inelastic pp collisions, and
$\kappa$ is the mean fraction of the kinetic energy $E_{\rm kin} =
E_p - m_p c^2$ of the proton transferred to the secondary meson
per collision. In a broad region from GeV to TeV energies, $\kappa
\sim 0.17$. In this approximation, then, an accurate knowledge of
the total inelastic cross section is needed to compute the
$\gamma$-ray emissivity.

Aharonian and Atoyan (2000) proposed that, since from the
threshold at $E_{\rm kin} \sim $ 0.3 GeV the cross section appears
to rise rapidly to about 30 mb at energies about $E_{\rm kin} \sim
$ 2 GeV, and since after that energy it increases only
logarithmically, a sufficiently good approximation is to assume
\ba \sigma & \sim & 30 \; (0.95 + 0.06 \ln (E_{\rm kin}/{\rm GeV}
) ) {\rm mb} \;\;\;\;\;\;{\rm for \;\;\; } E > 1 {\rm GeV}
\nonumber
\\ & \sim & 0 \;\;\;\;\;\; {\rm otherwise}. \label{AA} \ea It can
be seen (e.g., Dermer 1986a) that different parameterizations of
the cross section below 1 GeV do not noticeably affect the results
of $\gamma$-ray emissitivities, since most of the $\gamma$-rays
are generated by primary protons having more energy than a few
GeV, provided the spectrum of primaries is sufficiently broad.

\begin{figure*}[t]
\centering
\includegraphics[width=6cm,height=7cm]{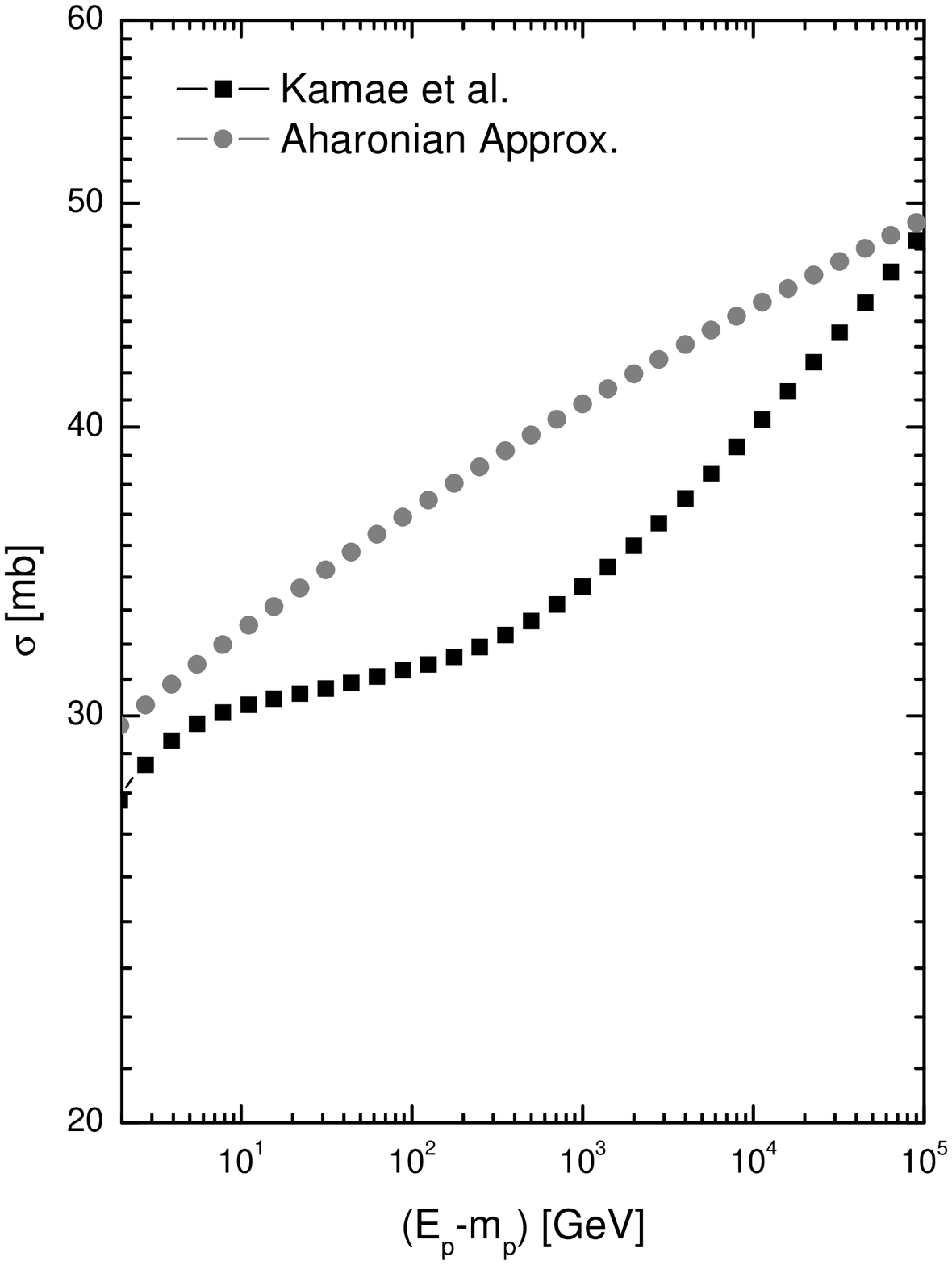}
\includegraphics[width=6cm,height=7cm]{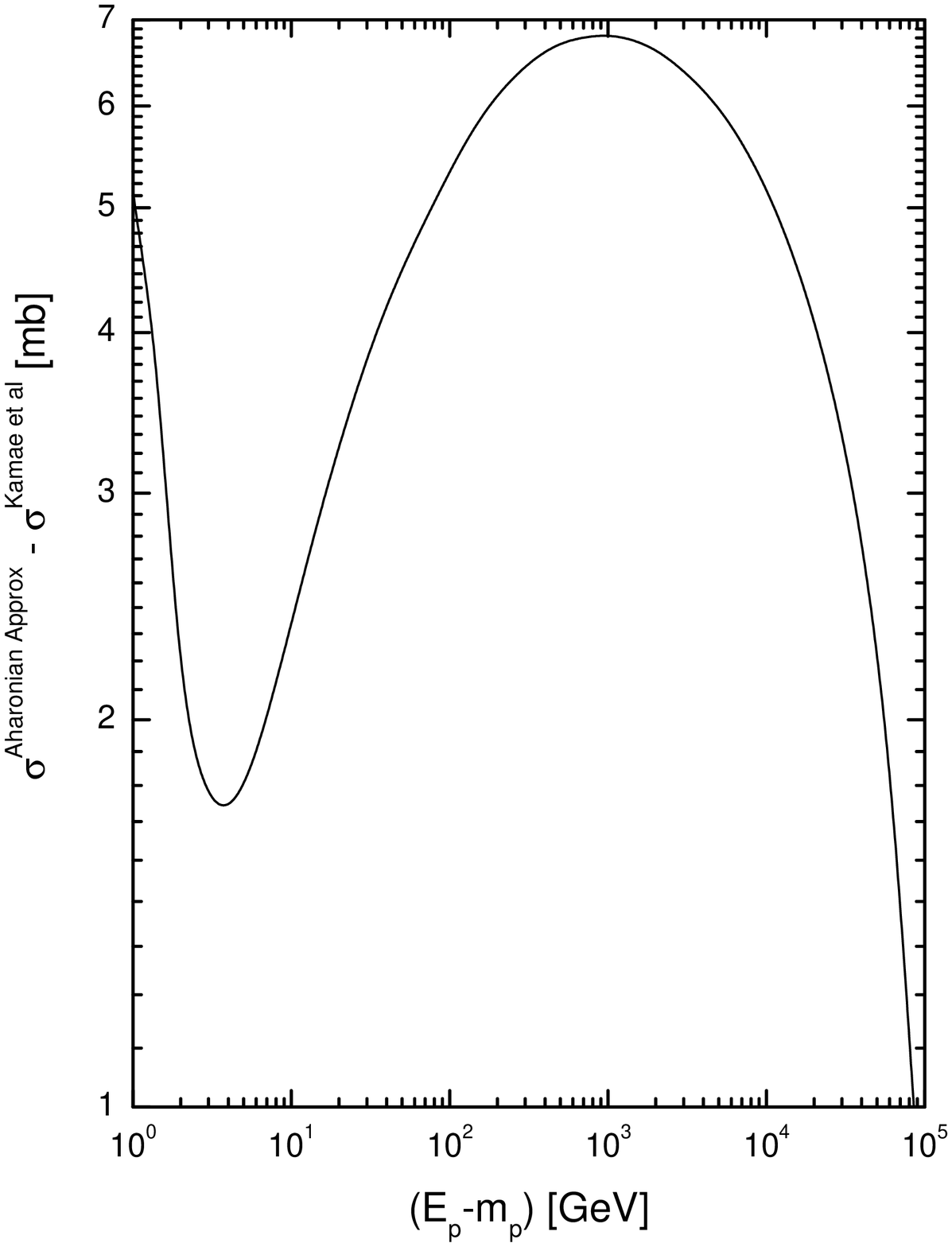}
\caption{Comparison between Kamae et al.'s model A, sum of
diffractive and non-diffractive contributions with Aharonian and
Atoyan's formula for the total inelastic cross section.}
\label{Kamae-Aha}
\end{figure*}

In a recent paper, Kamae et al. (2005) introduced the effect of
diffractive interactions and scaling violations in $pp \rightarrow
\pi^0$ interactions. The diffractive interactions contribution was
usually neglected in all other computations of $\gamma$-ray
emissivity from neutral pion decay to date, and thus one would
expect an increase in the predicted fluxes. Kamae et al.'s best
model, dubbed A, for the inelastic  (not inclusive)
cross section is given in
Table 1 of their paper, columns 2 and 3. When one compares the sum
of both diffractive and non-diffractive contributions of Kamae et
al.'s model with the Aharonian and Atoyan's formula, one sees that
the latter produces an actually larger (but quite close) cross
section. Figure \ref{Kamae-Aha} shows these results above proton kinetic
energies of 1 GeV, as well as the difference between these cross
sections. Kamae et al.'s model A was compared with Hagiwara's
(2002) compilation of pp cross section measurements and found in
good agreement. When multiplicity is taken into account, Kamae et
al.'s model also agrees with the data on inclusive cross sections,
a point we discuss in more detail below.

\begin{figure*}[t]
\centering
\includegraphics[width=6cm,height=7cm]{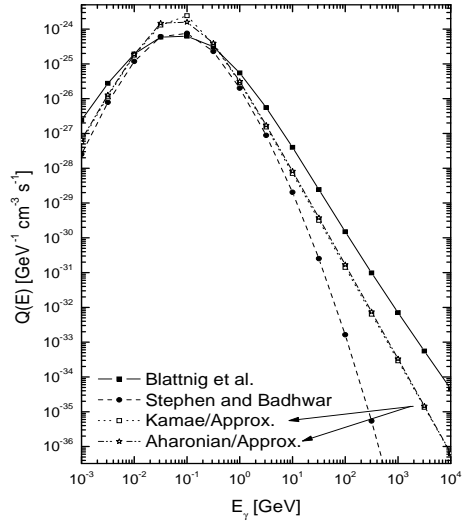}
 \caption{Comparing $\gamma$-ray emissivities within the
$\delta$-function approximation (Kamae et al.'s model A and
Aharonian and Atoyan's formula for the inelastic cross section)
with computations using differential cross section
parameterizations.
}
\label{All_crosssec}
\end{figure*}

Figure \ref{All_crosssec} shows a comparison of the $\gamma$-ray emissivity
obtained when using Kamae et al.'s model A and equation
(\ref{delta}). Curves are practically indistinguishable in this
scale, and their ratio is well within a factor of $\sim 1.3$. In
this comparison, the proton spectrum is the Earth-like one,
$J_p(E_p)=2.2\,E_p^{-2.75}$ protons cm$^{-2}$ s$^{-1}$ sr$^{-1}$
GeV$^{-1}$ and $n=1$ cm$^{-3}$. The resulting $\gamma$-ray
emissivity is multiplied by 1.45 to give account of the
contribution to the pion spectrum produced in interactions with
heavier nuclei both as targets and as projectiles (Dermer 1986a).
Aharonian and Atoyan's expression for the cross section produces a
slightly larger value of emissivity than that obtained with
Kamae's model A, including non-diffractive interactions.


\subsection*{Differential cross sections parameterizations}

Recently, Blattnig et al. (2000) developed parameterizations of
the differential cross sections.  They have presented a
parameterization of the Stephens and Badhwar's (1981) model by
numerically integrating the Lorentz-invariant differential cross
section (LIDCS). The expression of such parameterization is
divided into two regions, depending on the (laboratory frame)
proton energy (Blattnig et al. 2000, see their equations 23 and
24). Blattnig et al. have also developed an alternative
parameterization that has a much simpler analytical form. It is
given by \ba
   && \hspace{-1.cm} \frac{d\sigma(E_{\pi^0}, E_p)}{dE_{\rm \pi^0}}
    = e^A {\rm mb}\;      {\rm GeV}^{-1}
    \label{param_neutral}
     \\
     {\rm with} \nonumber \\
&&   \hspace{-1.cm} A={\left(
    -5.8-\frac{1.82}{(E_p-m_p)^{0.4}}
     +\frac{13.5}{(E_{\rm \pi^0}-m_{\rm \pi^0})^{0.2}}
     -\frac{4.5}{(E_{\rm \pi^0}-m_{\rm \pi^0})^{0.4}} \right)} \nonumber
     \ea
Both parameterizations were integrated and compared with
experimental results up to $\sim 50$ GeV in Blattnig et al.'s
(2000) paper. It was found that a single expression is needed to
represent the total inclusive cross section, \be \sigma_{\pi^0}
(E_p)= (0.007 +0.1 \frac{\ln (E_p-m_p)}{(E_p-m_p)} +
\frac{0.3}{(E_p-m_p)^2} )^{-1} {\rm mb}, \label{ITI} \ee where
rest masses and energies must be given in units of GeV.
These two differential cross section parameterizations proposed by
Blattnig et al. are not deprived of problems if extrapolated to
high energy. The parameterization of the Stephen and Badhwar's
model grossly underpredicts, whereas the newest Blattnig et al.
(equation \ref{param_neutral}) overpredicts, the highest energy
pion yield. The $\gamma$-ray photon yield that is output of the
use of these two differential cross sections in Equations
(\ref{1},\ref{pion-prog}) is also shown, for an Earth like
spectrum, in Figure \ref{All_crosssec}.

\begin{figure*}
\centering
\includegraphics[width=6cm,height=7cm]{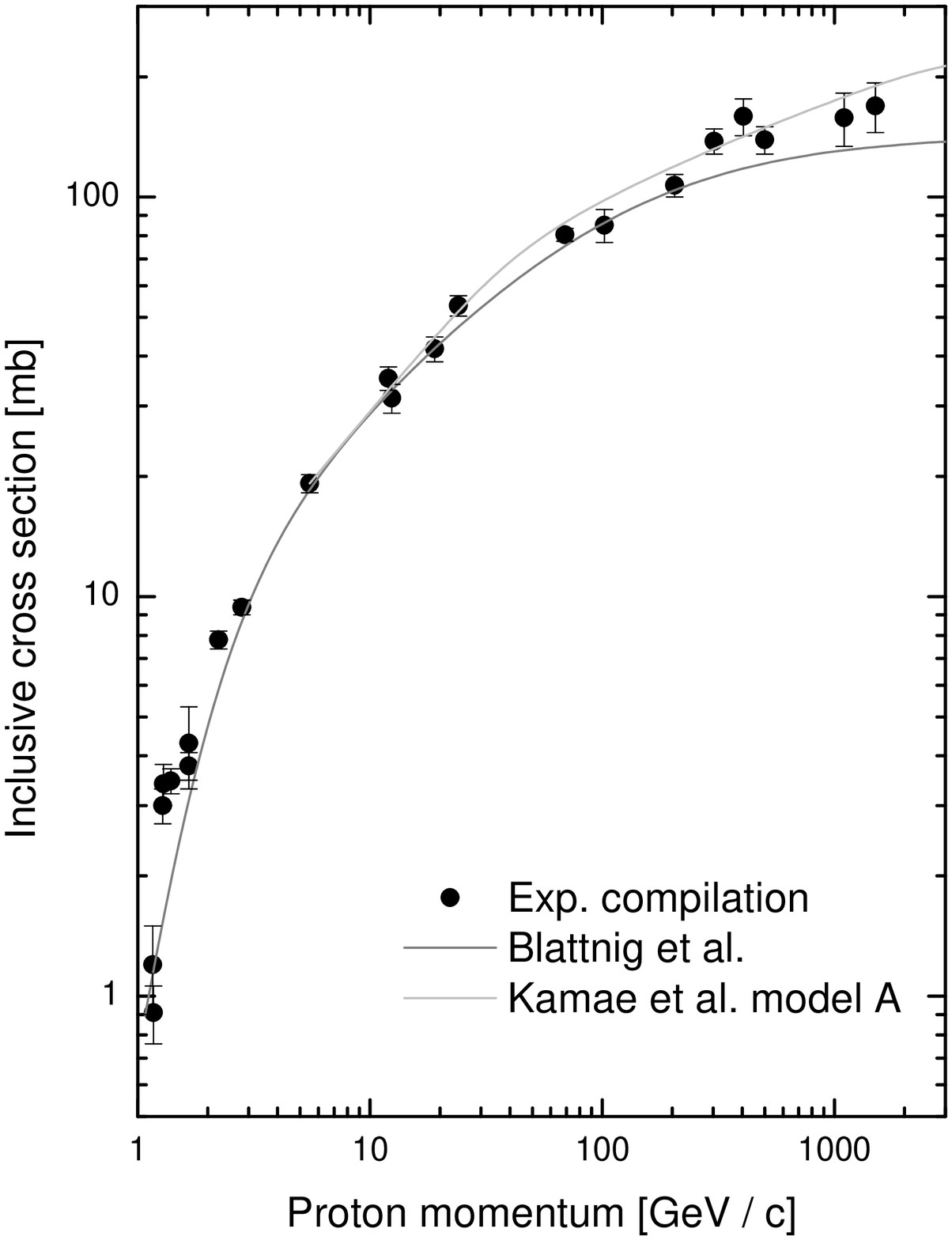}
\includegraphics[width=6cm,height=7cm]{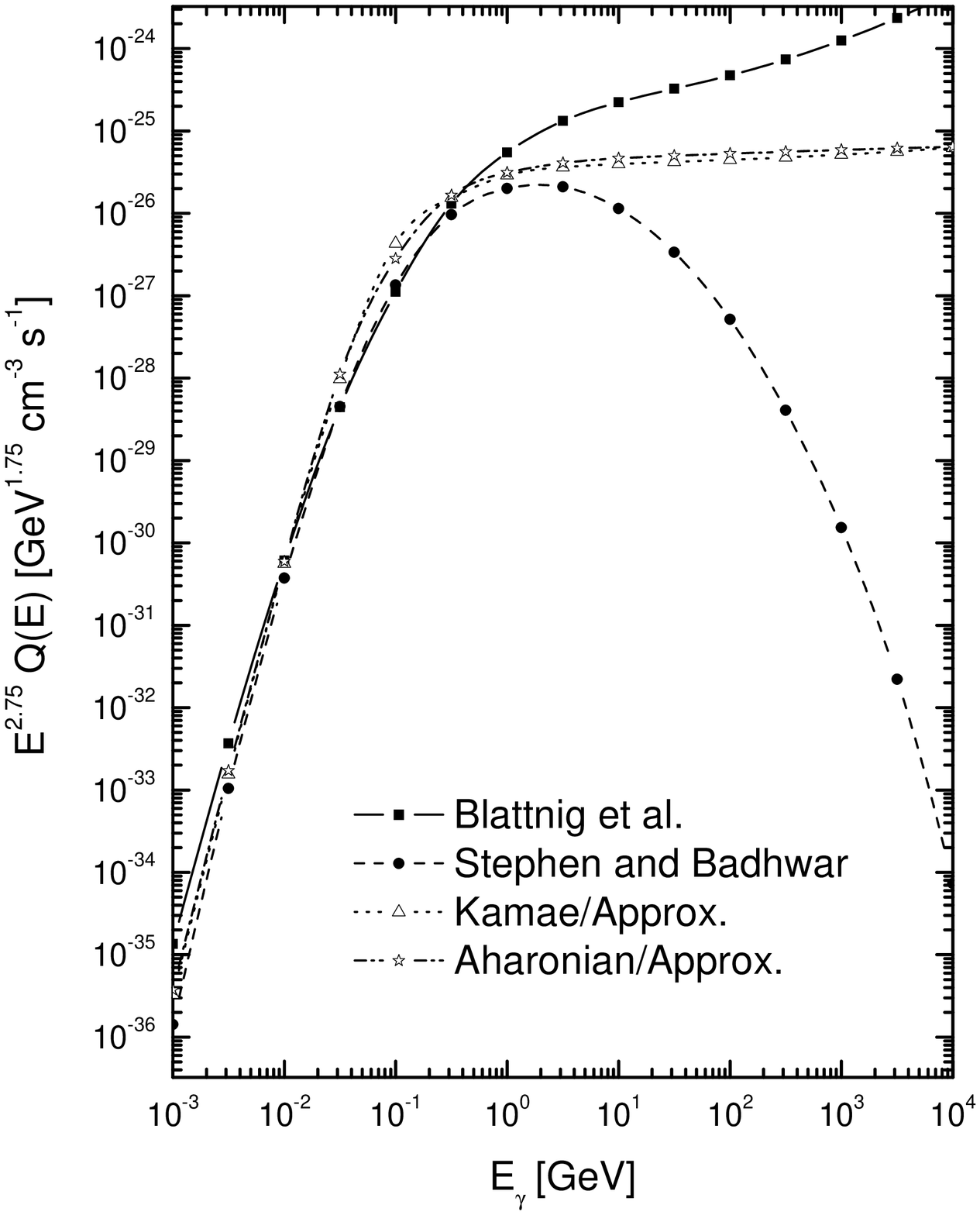}
\caption{Left: Comparing inclusive cross sections. Kamae et al.'s
model A data come from their figure 5. Blattnig et al. curve is
obtained from Equation (\ref{ITI}) and experimental compilation is
from Dermer (1986b). Right: $\gamma$-ray emissivities as in Figure
\ref{All_crosssec} multiplied by $E^{2.75}$, with 2.75 being the slope of the
proton primary spectrum. }
\label{emissiv}
\end{figure*}

However, the inclusive total inelastic cross section (\ref{ITI})
seems to work well at energies higher than 50 GeV, what we show in
Figure \ref{emissiv} together with a compilation of
experimental data (Dermer 1986b). 
We also show in the same Figure the results for the inclusive
total cross section from model A of Kamae et al. (2005), obtained
from his figure 5. Indeed, Kamae et al.'s model produces a
slightly higher cross section, although both correlate reasonably
well with experimental data, at least up to 3 TeV.\footnote{This
figure enlarge the comparison of Blattnig et al. (2000) inclusive
cross section with experimental data (see their figure 4), where
only three low-energy data points from Whitmore (1974) were
considered.} It is the differential cross section parameterization
(given by Equation \ref{param_neutral}) the one that looks
suspicious at such high energies. Figure \ref{emissiv}, right panel,
shows the emissivities as computed in the different approaches multiplied
by $E^{2.75}$. As expected, Stephen and Badhwar's parameterization
falls quickly at high energy whereas both Kamae et al.'s and
Aharonian and Atoyan's cross sections secure that the
$\gamma$-rays emitted maintain an spectrum close to that of the
proton primaries. For $\gamma$-rays above few TeV, i.e.,
$\gamma$-rays mostly generated by protons above few tens of TeV,
Blattnig et al. differential cross section parameterization makes
the $\gamma$-ray emitted spectrum much harder than the proton
spectrum that produced them. This signals that a direct
extrapolation of Eq. (\ref{param_neutral}) for computing photon
emissivity above TeV with Eq. (\ref{pion-prog}) induces
overpredictions of fluxes.

\begin{table*}
\begin{center}
\centering \caption{Integrated emissivities for an Earth-like
spectrum. Values are in units of photons  cm$^{-3}$ s$^{-1}$.}
\begin{tabular}{l c c c }
\hline
Parameterization/Approx. & $E>100$ Mev & $E>100$ GeV  & $E>315$ GeV\\
\hline
Blattnig et al.     &    3.2E-25      &        2.2E-28 & 4.7E-29\\
Kamae et al.        &    4.8E-25      &        1.8E-29 & 2.6E-30\\
Aharonian           &    4.0E-25      &        2.2E-29 & 3.1E-30\\
Stephen and Badwhar (from Blattnig et al.) &    2.2E-25      &        1.8E-31 & 1.9E-33\\
\hline
\end{tabular}
\label{integr}
\end{center}
\end{table*}

Table \ref{integr} presents the results for the integrated
emissivity, $\int_E \; Q(E) \,dE$, with $Q(E)$ being the different
curves of Figure \ref{All_crosssec}. To obtain integrated fluxes from a source of
volume $V$ at a distance $D$ one has to multiply by the constant
$V/(4\pi D^2)$, so that the difference in integrated emissivities
indeed represent those among integrated fluxes. As Table \ref{integr}
shows --disregarding those coming from the Blattnig et al.'s
parameterization of Stephen and Badwhar's results which are quoted
here just for completeness-- above 100 MeV, differences are less
than a factor of 1.5, which most likely is washed away by other
uncertainties in any given model. But above 300 GeV, difference
are larger and a conservative choice is in order.

If interested in the GLAST-domain (say, $E>100$ MeV, $E<50$ GeV)
predictions, the most conservative choice seems to be the use of
the Blattnig et al. new differential cross section
parameterization (Eq. \ref{param_neutral}), with no other
approximation, in Equations (\ref{1}) and (\ref{pion-prog}). This
choice, while not taking into account non-diffractive processes
will possibly {\it slightly underpredict} the integrated flux (as
shown in Table  \ref{integr}). Up to this moment, there is no public
parameterization of the differential cross section including
diffractive effects, but one is to be presented soon (T. Kamae,
private communication). By using Blattnig et al. approach, there
is no $\delta$-function approximation involved nor an ad-hoc
histogram of particle numbers as proposed in the treatment of
Kamae et al. (2005). One has an analytical expression that can be
directly used in the numerical estimates of Eq. (\ref{pion-prog}).
However, the price to pay is that this form of computation cannot
be considered reliable at higher energies and should not be used.

For the IACTs-domain ($E>100$ GeV) the safest and also
computationally-preferable choice appears to be to take either
Kamae et al.'s model A, or even the simpler Aharonian and Atoyan's
expression (Eq. \ref{AA}) and a $\delta$-function approximation.
This approach would probably be slightly underestimating the
integrated flux at such high energies. All in all, assuming either
Kamae et al.'s or Aharonian and Atoyan's expression for all
energies does not import substantial differences, as Table
\ref{integr} shows, and is computationally preferable.

\begin{figure*}
\centering
\includegraphics[width=6cm,height=7cm]{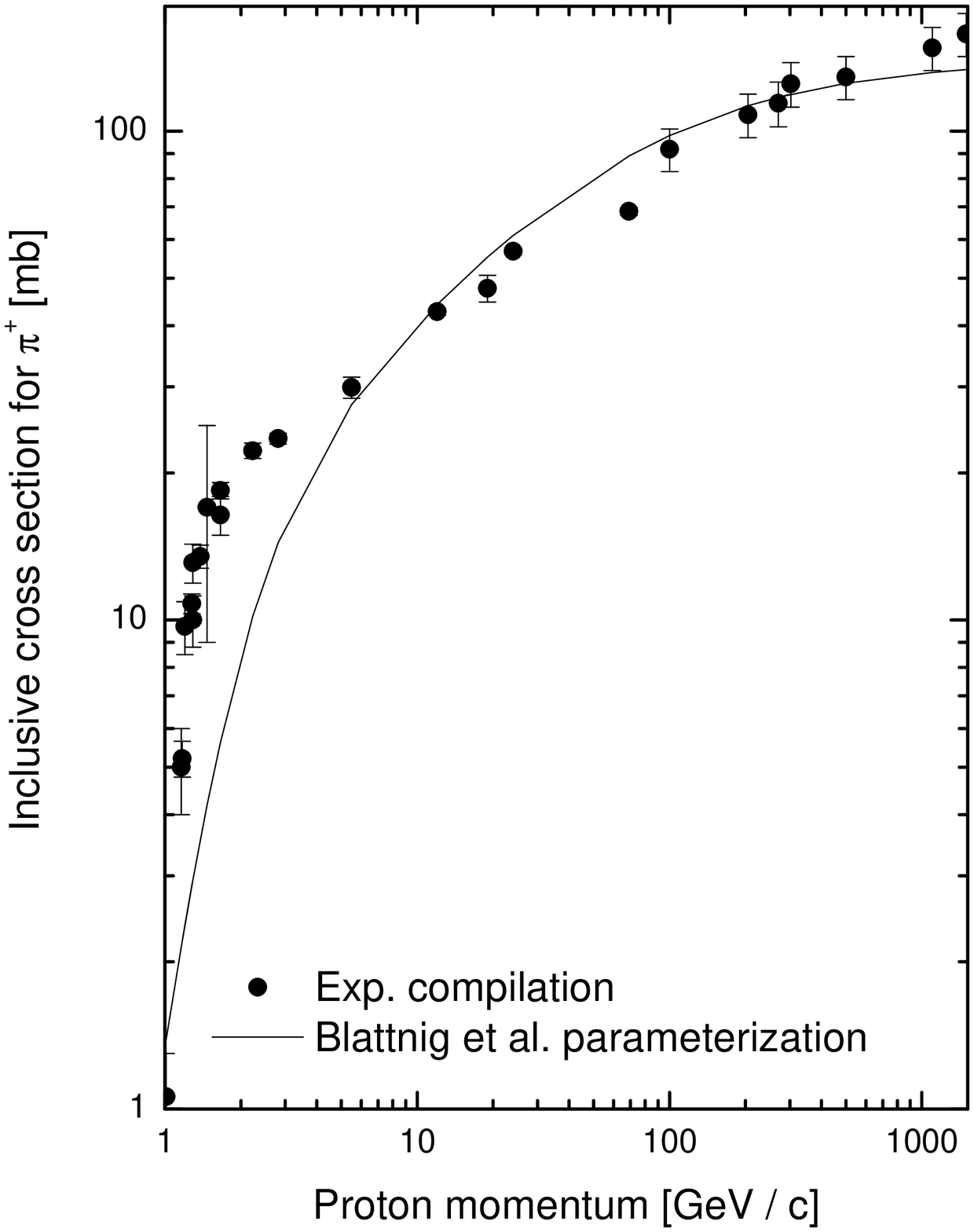}
\includegraphics[width=6cm,height=7cm]{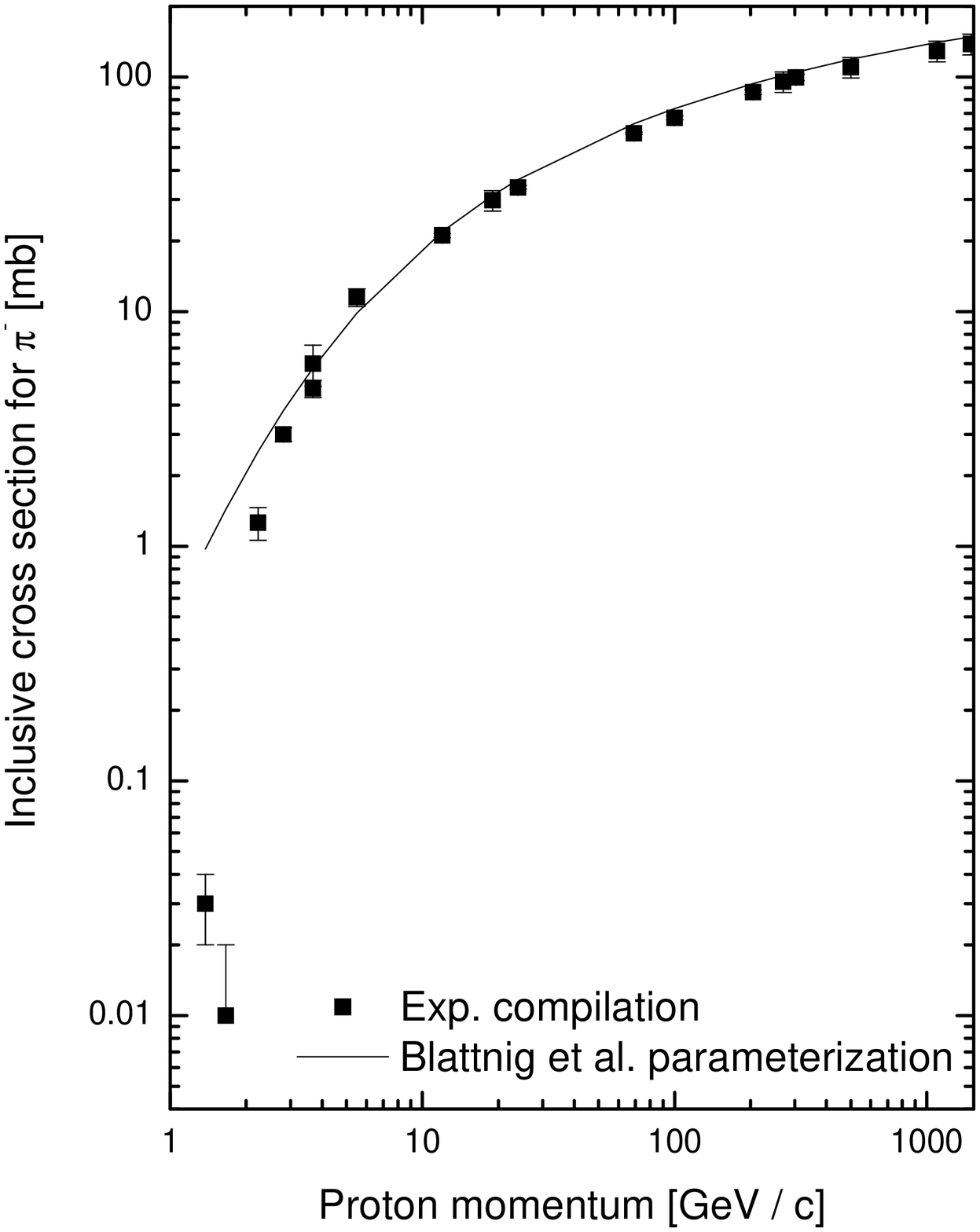}
\caption{Comparing inclusive cross sections for charged pions. Solid
curves are obtained from Equations (30) and (31) of  Blattnig et
al.'s work (2000b), and experimental compilation is from Dermer
(1986b).  }
\label{inclusive_charged}
\end{figure*}

Finally, in Figure \ref{inclusive_charged} we compare the inclusive
cross sections for charged pions with experimental data up to about 1 TeV,
which are again found in agreement with experimental data (except for a
bunch of data points at low proton energies, in the case of
positive charged pions). In any case, for situations where the
density of cosmic rays or of target nuclei or both are high,
neutral pion decay is expected to be the dominant process above
100 MeV, so that possible uncertainties in parameterizations of
charged pions cross sections are not expected to play a
significant role in the prediction of fluxes.

\section*{Acknowledgments}

The work of ED-S was done under a FPI grant of the Ministry of
Science and Tecnology of Spain. The work of DFT was performed
under the auspices of the U.S. D.O.E. (NNSA), by the University of
California Lawrence Livermore National Laboratory under contract
No. W-7405-Eng-48. We thank Juan Cortina, Igor Moskalenko, and
Olaf Reimer for comments. We thank the referee, V. Dogiel, for
remarks that led to an improvement of this work.

\end{document}